# A history of violence: insights into post-accretionary heating in carbonaceous chondrites from volatile element abundances, Zn isotopes, and water contents

Brandon Mahan [a,*], Frédéric Moynier [a,b], Pierre Beck [b,c], Emily A. Pringle [a], Julien Siebert [a,b]

[a] *Institut de Physique du Globe de Paris, Université Paris Diderot, Université Sorbonne Paris Cité,*

*CNRS UMR 7154, 1 rue Jussieu, 75238 Paris Cedex 05*

[b] *Institut Universitaire de France, Paris, France*

[c] *UJF-Grenoble 1, CNRS-INSU, Institut de Planétologie et d'Astrophysique de Grenoble (IPAG), UMR 5274, Grenoble F-38041, France*

*corresponding author, email: mahan@ipgp.fr, phone: +33 (07) 62 93 56 12*

## Abstract

Carbonaceous chondrites (CCs) may have been the carriers of water, volatile and moderately volatile elements to Earth. Investigating the abundances of these elements, their relative volatility, and isotopes of state-change tracer elements such as Zn, and linking these observations to water contents, provide vital information on the processes that govern the abundances and isotopic signatures of these species in CCs and other planetary bodies. Here we report Zn isotopic data for 28 CCs (20 CM, 6 CR, 1 C2-ung, and 1 CV3), as well as trace element data for Zn, In, Sn, Tl, Pb, and Bi in 16 samples (8 CM, 6 CR, 1 C2-ung, and 1 CV3), that display a range of elemental abundances from case-normative to intensely depleted. We use these data, water content data from literature and Zn isotopes to investigate volatile depletions and to discern between closed and open system heating. Trace element data have been used to construct relative volatility scales among the elements for the CM and CR chondrites. From least volatile to most, the scale in CM chondrites is Pb-Sn-Bi-In-Zn-Tl, and for CR chondrites it is Tl-Zn-Sn-Pb-Bi-In. These observations suggest that heated CM and CR chondrites underwent volatile loss under different conditions to one another and to that of the solar nebula, e.g. differing oxygen fugacities. Furthermore, the most water and volatile depleted samples are highly enriched in the heavy isotopes of Zn. Taken together, these lines of evidence strongly indicate that heated CM and CR chondrites incurred open system heating, stripping them of water and volatiles concomitantly, during post-accretionary shock impact(s).

# 1. Introduction

Carbonaceous chondrites (CCs) have long been recognized as early solar nebula remnants (e.g. Anders 1964; Palme et al. 2014), as their elemental abundances reveal striking similarities to that of the Sun's photosphere. These include the volatile elements, with 50% condensation temperatures under nebular condition ($T_{50}$) between 180K and 665K, and the moderately volatile elements, 665K< $T_{50}$ <1135K ($T_{50}$ from Lodders, 2003), which act as measuring devices for the conditions of various planetary processes - accretion, differentiation, outgassing, etc. - due to their thermodynamic sensitivity (e.g. T, P, $fO_2$ or oxygen fugacity). For ease of reading, these elements are hereafter collectively referred to simply as volatile. Additionally, similarities between the bulk hydrogen and nitrogen isotopic signatures of CCs and Earth imply that CCs may have supplied Earth with its inventory of water (e.g. Alexander et al., 2012). Linking observations between volatile element abundances, water content, and isotopic systems in CCs will further propel our understanding of thermal alteration and volatilization.

Carbonaceous chondrites display depletions in volatile elements, in descending order of volatile element abundances from CI (Ivuna-type) to CV (Vigarano-type) (O'Neill and Palme 1998; Palme and O'Neill, 2014). These characteristic depletion trends have been attributed to various processes of volatility controlled elemental fractionation, including but not limited to: incomplete condensation in the early solar nebula during chondrite formation (e.g. Wasson and Chou 1974; O'Neill and Palme, 2008), two-component mixing (e.g. Anders, 1964), X-wind controlled formation (Shu et al., 1996), thermal processing in the solar system's parent molecular cloud (e.g. Huss et al., 2003), and/or inheritance from the interstellar medium (Yin, 2005). A succinct overview of these multiple working hypotheses, and references to their progenitors can be found in Bland et al. (2005). Whatever the process, or combination of processes, each CC type was endowed with a distinct initial signature of volatile element abundances (e.g. Xiao and Lipschutz, 1992; Lodders, 2003; Bland et al., 2005; Weisberg et al., 2006; Choe et al., 2010; Palme and O'Neill, 2014). After formation of the various chondrite components and emplacement of these primary features, secondary processes such as thermal and aqueous alteration began modifying their mineralogic and chemical composition (Huss et al., 2006), and the extent of alteration due to these processes has been integral to classification schemes of chondrites (e.g. van Schmus and Wood, 1967; Weisberg et al., 2006).

Historically, thermal alteration has been most often observed and most extensively studied in the ordinary chondrites (OCs), which display a range from un-metamorphosed (petrologic type 3) to highly metamorphosed (petrologic type 6) (Weisberg et al., 2006), wherein a higher degree of metamorphism lends also to a slight depletion in volatile elements (e.g. Zn, In, Cd, Bi, Tl) (Huss et al., 2006; Schaefer and Fegley 2010a). Thermal metamorphism in CCs is generally less pervasive and severe, with most having a petrologic type ≤3 (e.g. Weisberg et al., 2006). A clear exception to this are the CK chondrites (Karoonda-type), which display petrologic types 3-6, with a paucity in type 3, and depletions in volatile elements greater than those of the CV (Weisberg et al., 2006; Huss et al., 2006), making them less than ideal for a systematic study of volatile loss and isotopic fractionation (due to low concentrations) in CCs. Within the CM (Mighei-type) and CR (Renazzo-type) chondrites however, there are plentiful normative samples as well as samples that have been altered by thermal events, thanks largely to Antarctic missions garnering some 22,000+ samples (according to Johnson Space Center). These CM and CR chondrites provide an ideal sampling pool for investigating post-formation volatility loss, as a strong case-normative baseline can be established for both groups, and thermal events have been indicated for samples in both groups (e.g. Akai, 1992; Tonui et al., 2003, 2014; Wang and Lipschutz, 1998; Abreu and Bullock, 2013; Alexander et al., 2012, 2013; Beck et al., 2014; Schrader et al., 2015), concomitant with severe loss of volatile elements (though still remaining in measurable amounts) (Paul and Lipschutz, 1989; Xiao and Lipschutz, 1992; Wang and Lipschutz, 1998). Moreover, it has been posited that the heat source for such thermally altered CCs - at least in the case of the CM chondrites - may be different than that of the OCs, i.e. the prevalent heat source for many CCs was late in its history and of short duration compared to OCs, whose heating occurred earlier and for longer durations (e.g. Nakamura, 2006; Nakamura et al, 2006). Further investigation of heated CCs is thus necessary to differentiate between potential heat source(s).

Aqueous alteration in carbonaceous chondrites ranges from essentially unaltered samples (petrologic type 3) to near complete hydration of silicates (petrologic type 1) (Weisberg et al., 2006). As a process, aqueous alteration is generally thought to have preceded thermal alteration in many CCs (e.g. Nakamura, 2005; Cloutis et al., 2012; Tonui et al., 2014), possibly even beginning pre-accretion (Bischoff, 1998), and occurred at temperature conditions less than ~300˚C (down to ~0˚C) (e.g. Brearley, 2006). This hydration, in terms of the abundance of water, is most prevalent in the CI chondrites (~20 wt% water, petrologic type 1), with decreasing amounts in the CM (~9 wt%) and CR (<9 wt%) chondrites, most of which are of petrologic type 2 (Brearley, 2006; Garenne et al., 2014). Aqueous alteration is quite ubiquitous in CCs and its extent, conditions and timing are varied, though this does not appear to have any first order effects on volatile element concentrations (e.g. Rubin et al., 2007 for CM chondrites). Moreover, terrestrial weathering, which may contribute to the overall water budget of meteorites, shows no clear relationship between weathering grade and alteration products (phyllosilicates/[oxy]hydroxides) (Cloutis et al., 2012; Garenne et al., 2014). However, the presence and abundance of water on the parent body(ies) of CCs can have a strong effect on the redox conditions prevalent during a subsequent heating event (Schaefer and Fegley, 2010b), and therefore may play an important (though indirect) role by changing the volatility of elements during volatilization (O'Neill and Palme, 2008).

Water, in any of its variously bound states ($H_2O$, oxy-hydroxide minerals, hydroxyl groups; Garenne et al., 2014), may be liberated from a meteorite parent body via volatilization in the event of heating. In this scenario, the initial hydration of the chondrite can be inferred from petrologic evidence such as the presence and abundance of phyllosilicates (Garenne et al., 2014; Howard et al., 2015). Subsequent thermal alteration can then be inferred by the reversion of these phyllosilicates to anhydrous phases (e.g. Akai, 1992), the oxidation of iron (Beck et al., 2012), and through spectral reflectance properties (e.g. Cloutis et al., 2011, 2012). In addition to petrologic-based evidence for heating events, $\delta D$ and H abundance analyses have been successfully employed to identify 'heated' specimens, specifically within the CM chondrite group (Alexander et al. 2012, 2013). In such manners, both the aqueous and thermal histories of heated meteorites can be discerned, and indeed have been documented in the literature for some CM and CR chondrites (e.g. Wang and Lipschutz, 1998; Alexander et al., 2012, 2013; Garenne et al., 2014). Furthermore, volatile loss has also been simulated experimentally in CM chondrites by tracking the depletions of volatile elements (e.g. Zn, In, Bi, Tl, and Cd) as a function of temperature in controlled heating experiments of the Murchison (CM2) meteorite in a low pressure (~$10^{-5}$ atm, or ~10 Pa) $H_2$-rich environment analogous to the solar nebula (i.e. reducing conditions), further indicating that volatilization upon heating can create distinct depletion trends in meteorites, however natural observations are not always in sync with such experimentally derived volatility sequences and depletions cannot be solely explained by the magnitude of heat (Matza and Lipschutz, 1977; Wang and Lipschutz, 1998; Nakamura, 2005). As the redox conditions prevailing during post-formation thermal alteration would affect the order of volatilization of the elements (e.g. O'Neill and Palme, 2008), condensation temperatures under solar nebula conditions (e.g. Lodders, 2003), as well as experiments in an $H_2$-rich environment (Matza and Lipschutz, 1977), are not strictly applicable. Discrepancies such as these, as well as the discrepancies in estimates of conditions such as heating temperatures (e.g. Table 5 in Cloutis et al., 2012) clearly define a need for future work.

Having such rich and varied literature regarding the CM and CR chondrites renders them ideal chondrite groups to use for investigating post-accretionary heating. By investigating multiple samples from these chondrite groups, we can explore the effects of heating on the water and volatile contents of carbonaceous chondrites, as well as the relative volatility of the different elements under conditions relevant to post-accretionary heating on CC parent bodies. Additionally, Zn isotopic composition has proven a very powerful tool for investigating evaporation processes, as Zn isotopes are strongly fractionated during volatilization, wherein preferential loss of the light isotopes to the gas phase occurs (e.g. Moynier et al., 2017). In this way, investigating the Zn isotopic signatures of CCs can provide further information on the conditions controlling volatile content variations, i.e. whether or not evaporative (or other) processes in an open system are responsible for the volatile element depletion, which should be evidenced by isotopically heavy residues (Day and Moynier, 2014). Moreover, these observations can further constrain the heating mechanism in CCs, its source(s) and timing, and thus provide substantial insights that compliment studies of organics, opaque mineral assemblages, and pre-solar grain abundances in CCs. Developing relative volatility trends for volatile elements in carbonaceous chondrites based on natural observations, and linking these observations to water contents and isotopic systems that act as state-change tracers, such as Zn, can provide a strong foundation upon which we can build a much richer and more complete comprehension of post-accretionary heating of CCs, honing our understanding of the thermodynamic conditions (P, T, $fO_2$) that created their volatile element abundance trends, as well as the heating mechanism(s) behind them.

Here we have analyzed the Zn isotopic composition and volatile element contents for a suite of carbonaceous chondrites, including case-normative samples and those indicated as having incurred heating events, with a focus on the CM and CR chondrites. We have combined elemental data with literature data for trace elements (Xiao and Lipschutz, 1992; Paul and Lipschutz, 1989; Zolensky et al., 1992, 1997; Wang and Lipschutz, 1998) and water content (Alexander et al., 2012; Garenne et al., 2014) to determine the relative volatility of the investigated elements in the CM and CR chondrites, and to link depletions in volatile elements with variations in water content in CCs. Additionally, through a comparison of elemental concentrations, water content and Zn isotopic data, we have provided robust evidence for open-system volatilization of Zn, and thus presumably water and other species in meteorites (and/or their parent bodies) that have undergone heating above ~700°C through post-accretionary impact heating.

## 2. Samples and Analytical Methods
### 2.1 Sampling pool

For this study, carbonaceous chondrites from the CM and CR groups (as well as 1 CV3 and 1 ungrouped C2) were analyzed to provide a broad contextual range, including case-normative samples to provide a strong baseline, as well as samples with evidence of thermal events. CM and CR chondrites were the primary focus as there are abundant samples, including samples from both the CM and CR groups displaying evidence of heating events (e.g. Alexander et al., 2012, 2013; Abreu, 2011; Abreu and Bullock, 2013; Beck et al., 2014; Tonui et al., 2014; Schrader et al., 2015), and therefore are of particular interest in regards to water content, volatile element concentrations and Zn isotopic fractionation. Approximately one half (13) of samples in the current study are demarcated as heated meteorites, many on the basis of H, C, and N abundances as evidenced in Alexander et al. (2012, 2013) and/or water contents as evidenced in Garenne et al. (2014). These include the CM chondrites ALH 84033, DOM 03183 (possibly heated), EET 87522, EET 96029, MAC 88100, MIL 05152 and MIL 07700. The two CM chondrites, PCA 02010 and PCA 02012, were included based on severe depletions in water content and petrologic indications of heating (Alexander et al., 2012, 2013; Garenne et al., 2014), as well as petrologic indications of intensive heating to at least 900°C associated with shock impact for PCA 02012 (Nakato et al., 2013). The CM-like ungrouped C2 (C2-ung) meteorite EET 83355 was included based on its identification as heated, the availability of water content data and because it has been reported both as an ungrouped C2 (Wang and Lipschutz, 1998; Garenne et al., 2014) and as a C2/CM2 (Xia and Lipschutz, 1992; Alexander et al., 2013), and as such shares petro-chemical (petrology/mineralogy) similarities with CM chondrites. The CR chondrites GRA 06100 and GRO 03116 were included based on petrographic indicators of shock heating (e.g. Abreu, 2011) and severe depletion in water contents (Garenne et al., 2014). The reduced CV3 RBT 04133 was also included for its available water content data and petro-chemical similarity with the CR chondrites, as it was initially classified as a CR chondrite (Weisberg et al., 2008), but has subsequently been reclassified as a reduced CV3 (Davidson et al., 2014); furthermore, it has been identified by Davidson et al. (2014) as heated, based on C and N isotopes, maturation of organics, and low pre-solar grain abundance. Zinc isotopic analysis was conducted on 20 CM chondrites and 6 CR chondrites (+1 CV3 and 1 ungrouped C2), and trace element analysis was conducted on 8 CM chondrites and 6 CR (+1 CV3 and 1 ungrouped C2) chondrites from this sampling pool.

**2.2 Dissolution and Chemical Purification**

Aliquots of ~50 mg for each sample were taken from larger powdered masses (approximately 1 g each) to ensure homogenized whole rock representation (e.g. Clayton and Mayeda, 1999; Schrader et al., 2011). All samples were dry-ground with mortar and pestle to a fine powder to facilitate chemical breakdown. Powdered samples were then dissolved in a concentrated $HNO_3$ / HF mixture to ensure full digestion. Sample solutions were then dried and the residuum re-dissolved in 6M HCl to dissolve fluoride compounds. After sequential digestion, the samples were again dried down and the residuum re-dissolved in 1.5M HBr in preparation for ion exchange chromatography. Isolation of Zn ions from other species was achieved using an anion exchange resin (Bio-Rad™ AG1 X8, 200-400 mesh) following the method used in a previous work (Mahan et al. 2017), and the technical details of which can be found elsewhere (Moynier et al., 2006; Moynier and LeBorgne, 2015). Prior to loading, the ion exchange resin was cleansed through repeated $HNO_3$ and Milli-Q water passes. Once loaded onto the exchange columns, Zn remains adsorbed to the resin as a bromine complex while other cationic species are eluted with HBr. Once elution of the matrix is completed, Zn is eluted using dilute (0.5M) $HNO_3$, then subsequently dried down in preparation for isotopic analysis.

**2.3 Zinc isotopic analysis (MC-ICPMS)**

All purified samples were dissolved in 0.1M $HNO_3$ for isotopic analysis. Zinc isotope mass spectrometry was performed on these purified solutions using the Thermo Scientific Neptune Plus MC-ICPMS (multi-collector inductively coupled plasma mass spectrometer) at the Institut de Physique du Globe de Paris (IPGP) following the method used in a previous work (Mahan et al. 2017), outfitted with an SSI quartz nebulizer / spray chamber (see Moynier and LeBorgne, 2015 for more details) and Faraday cups in conjunction with $10^{11}$ ohm resistors.

Corrections for instrumental mass bias were performed using sample-standard bracketing with the JMC Lyon Zn reference solution (Marechal et al., 1999). $\delta^n Zn$ for all samples have been reported relative to the JMC Lyon standard and calculated using conventional delta notation via Eq. 1,

Equation (1) $$\delta^n Zn = \left[ \frac{^nZn/^{64}Zn_{Sample}}{^nZn/^{64}Zn_{JMC\,Lyon}} - 1 \right] * 1000$$

where $n$ = 66 or 68. All Zn fractionation reported within this study occurred through mass-dependent processes, and therefore only $\delta^{66}Zn$ will be discussed hereafter in the text. Wherever possible, i.e. when sample concentrations were sufficiently high, Zn isotopic compositions were analyzed multiple times (at least 3 replicates for most samples). Errors for all isotopic analyses have been conventionally reported as two standard deviations (2σ). External reproducibility and accuracy of the entire methodology (dissolution, chemical purification, mass spectrometry) was rigorously evaluated concomitant with another study (Mahan et al., 2017), as samples and data for both projects were processed and acquired contemporaneously. In this evaluation, the 2σ error for the geostandard BHVO-2 was 0.02‰, and $\delta^{66}Zn$ values obtained for the geostandards BHVO-2 (0.30 ±0.02‰) and AGV-2 (0.25 ±0.08‰) were in good agreement with reference values of 0.28 ±0.04‰ and 0.29 ±0.03‰ (Moynier et al. 2017). Moreover, five complete replicates from samples containing Zn with different matrix materials yielded a 2σ error of 0.04‰. Taken together, these observations indicate high analytical precision and external reproducibility for the entire Zn isotope methodology. Isotopic data for all meteorites in the current study, along with water contents (Garenne et al., 2014), are detailed in Table 1.

**2.4 Trace element analysis**

After sequential digestion, aliquots of all samples were taken from the HBr solutions, dried down, then re-dissolved in 1% $HNO_3$ for trace element analysis with the Thermo Fisher Element 2 HR-ICPMS (high resolution inductively coupled plasma mass spectrometer) at IPGP. Sample solutions were diluted such that the elements of interest were in concentrations within the calibration range of the instrument. Calibration of the instrument was conducted prior to each session using a stock solution created from mixing aliquots of standard solutions with precisely known concentrations, and subsequently diluting this stock solution with 1% $HNO_3$. Aliquot masses for all standard solutions, along with the mass of the 1% $HNO_3$ dilutant, were directly measured so that precise concentrations of elements in the stock and calibration solutions could be calculated to ensure the most accurate data treatment possible. Calibration data and raw data generated by the mass spectrometer were then input into the open-access uFREASI (user-Friendly Elemental dAta procesSIng) software module (Tharaud et al., 2015) for calculation of trace element concentrations. Trace element concentrations for all samples can be found in Table 2.

**3. Results**
**3.1 Zinc isotopes**

Zinc isotope data for all samples have been reported in Table 1 along with petrologic type and weathering grade. Table 1 also reports water contents previously estimated by Garenne et al. (2014) or calculated stoichiometrically using H content from Alexander et al. (2012) and a 1/8 mass ratio of H to O in water. Error for all data are two times the standard deviations (2σ), and reproducibility for the entire method is 0.04‰ for $δ^{66}Zn$ as discussed in ***Section 2.3*** and Mahan et al. (2017). For most CM and CR chondrites, $δ^{66}Zn$ values do not vary significantly and are on par with literature values (e.g. Luck et al., 2005; Moynier et al., 2017; Pringle et al., 2017). $δ^{66}Zn$ values for CM chondrites as a function of water content can be found in Fig. 1. The heated CM chondrites PCA 02010 and PCA 02012 (5.42 ±0.44‰ and 16.73 ±0.09‰, respectively) and the heated CR chondrite GRO 03116 (4.77 ±0.03‰) (Alexander et al., 2012, 2013; Garenne et al., 2014 for CMs; Abreu, 2011; Schrader et al., 2015 for CRs), which are the most volatile and water depleted samples in their respective groups (all below 3 wt% $H_2O$), the $\delta^{66}Zn$ values show significant enrichment in the heavy Zn isotopes (Tables 1 and 2). The heated CR chondrite GRA 06100 has a significantly negative $\delta^{66}Zn$ value (-1.02 ±0.08‰) with no concomitant depletion in Zn.

### 3.2 Trace element contents

Concentration data from HR-ICPMS for the elements Zn, In, Sn, Tl, Pb and Bi for a suite of carbonaceous chondrites (CM and CR) are listed in Table 2, along with $\delta^{66}Zn$ and water contents (as $H_2O$ wt%, as in Table 1). Elemental abundances range from typical values in accord with reported averages in the literature (e.g. Lodders and Fegley, 1998), to intensely depleted in volatiles. Additionally, BHVO-2 was analyzed concomitant with other samples (Table 2) and the values determined are in good agreement with literature data (e.g. Barrat et al., 2012; Gale et al., 2013). Water contents have also been included in Table 2 for ease of reading. For comparison of relative abundances, CM and CR concentrations from this study along with literature data for Zn, Tl, Bi, and In (Table 3) have been normalized (via local maxima, see ***Section 4.1*** for more details) and plotted against Zn as a reference element in Fig. 2 for CM chondrites and in Fig. 3 for CR chondrites. Normalized element concentrations for the type 2 CM chondrites have been plotted against water contents in Fig. 4 in order to investigate volatile element depletion as a function of water content. Elemental abundances for CM chondrites in Fig. 4 display normative values to a threshold water content ($H_2O$ wt%) in the sample of ~5 wt%, below which the volatile element abundances become significantly depleted, specifically for carbonaceous chondrites reported to have undergone extensive heating, e.g. the CM chondrites PCA 02010 and PCA 02012 (Alexander et al., 2012; Garenne et al., 2014) (Table 2, Fig. 4). The CR chondrite GRO 03116, also reported to have undergone a heating event (Abreu, 2011; Alexander et al., 2013; Briani et al., 2013), likewise displays intensely depleted volatile element abundances concomitant with low water contents (Table 2). The heated CR chondrite GRA 06100 (e.g. Abreu, 2011) displays an anomalous and significantly negative $\delta^{66}Zn$ value with no concomitant depletion in Zn abundance (63 ±9ppm), and equally vexingly displays an anomalously high Pb concentration (7.58 ±2.93ppm). The most volatile-rich CR chondrite investigated, GRO 95577, has been previously classified as a CR1 chondrite due to its extensive aqueous alteration, and does not display depletions in chalcophile elements characteristic of CRs (Weisberg and Huber, 2007). Trace element data from this study for GRO 95577 agree with the observations of Weisberg and Huber (2007), wherein the abundances of the volatile elements approximate ~60% CI.

## 4. Discussion
### 4.1 Volatility in CM and CR chondrites

The canonical scale of volatility for the elements is defined by the temperature at which 50% of an element is condensed into the solid phases under solar nebula conditions ($T_{50}$), i.e. in a highly reducing $H_2$ environment with a log$fO_2$ of approximately -20 (or IW-7) at low pressure (~10 Pa) (e.g. Lodders, 2003). However, it is well known that the volatility of an element depends on the speciation of the element in the gas and in the solids and therefore it is strongly dependent on redox conditions. This factor may then largely contribute to the re-arrangement of element volatility trends in CCs in comparison to that of Lodders (2003). These changes in redox state have been attributed to the fact that post-accretionary, post-hydration heating conditions occur under different conditions than nebular condensation, i.e. in more hydrated and therefore more oxidizing conditions and most likely at varying temperature and pressure conditions than that of the solar nebula (e.g. O'Neill and Palme, 2008). Therefore, to better understand the depletions of volatile elements observed in carbonaceous chondrites, it is vital to characterize the volatility of these elements in an appropriate reference frame. Previous experimental work (Matza and Lipschutz, 1977) has determined the relative volatility (or thermal mobility) of a suite of elements in the Murchison CM2 meteorite via heating experiments carried out in a low-pressure environment (~10 Pa), and literature data for heated meteorites are in general agreement with this depletion trend (e.g. Matza and Lipschutz, 1977; Wang and Lipschutz, 1998). However, observations on natural samples have not been investigated as a complimentary means of determining the relative volatility of elements in carbonaceous chondrites (though a similar approach has been taken for ordinary chondrites, see Schaefer and Fegley, 2010a).

By investigating the volatile element contents in CM and CR chondrites, including case-normative samples as well as samples that have undergone heating events, it is possible to construct relative volatility scales for these meteorite groups. These can be determined by first normalizing concentration data for all samples, wherein concentrations for each element are normalized to a local maximum, i.e. the largest value for a given element within each respective dataset, then plotting these normalized values relative to a reference element (Zn in the current study). Data for each element determined through HR-ICPMS in the current study were normalized to their respective local maxima. Likewise, data for each element from the literature, determined via neutron activation analysis (NAA), were normalized to local maxima for the literature dataset. Normalization by this approach was used as it removes methodological artifacts (i.e. differences in acquisition media, calibration, etc.) and more importantly because via such normalization all data are given the same index, and therefore the slope calculated by plotting each element against Zn (with Zn on the x-axis) denotes the volatility of that element relative to Zn, wherein the largest slope indicates that the element is the most volatile, and vice versa. Plots of M/Zn (where M is the other element) for CM chondrites can be found in Fig. 2, and values for the slope, y-intercept and coefficient of determination ($R^2$) for each element pair can be found therein along with corresponding errors (1σ) in parentheses. Where available, data from literature (Xiao and Lipschutz, 1992; Paul and Lipschutz, 1989; Zolensky et al., 1992, 1997; Wang and Lipschutz, 1998) have been included to provide the most robust assessment possible (see Table 3). Using this approach, the relative volatility in CM chondrites for the six elements investigated - from least volatile to most – has been determined as Pb-Sn-Bi-In-Zn-Tl. It should be noted that this is the most likely order of volatility for the CM chondrites, however as the variation in slope is small relative to the uncertainties, different trends cannot be ruled out at this time. The same approach has been applied to the CR chondrites in the current study (Fig. 3), yielding a relative volatility scale of Tl-Zn-Sn-Pb-Bi-In (see Table 4). Figures 2f and 3f ("CM gradient" and "CR gradient") display the slopes for all elements as a function of their respective y-intercept. The relationship between the slope and y-intercept is such that as the volatility of a given element decreases relative to Zn, the amount of that element remaining after all Zn has been exhausted increases (e.g. Sn and Tl in Fig. 3).

The relative volatility scales for both CM and CR chondrites determined via this method are different than the volatility trend for solar nebula conditions (Table 4), which may indicate a change in conditions, such as an increase in $fO_2$ (O'Neill and Palme, 2008), illustrating the necessity of constructing relative volatility scales for the elements which inherently consider the conditions and myriad processes that produce the abundance trends we observe in heated carbonaceous chondrites. Once established, these relative volatility scales can provide important information about the conditions of post-accretionary heating such as temperature (e.g. Wang and Lipschutz, 1998), $fO_2$ (e.g. O'Neill and Palme, 2008), and pressure (Schaefer and Fegley (2010a) for ordinary chondrites) for volatile elements. For example, the conditions of element mobilization (e.g. via volatilization) in the CM and CR chondrites were likely more oxidizing than that of their formation in the solar nebula (e.g. Lange and Ahrens, 1982; Schaefer and Fegley, 2010b). Under the more reducing conditions of the solar nebula, elements such as Zn and In become more volatile (O'Neill and Palme, 2008), a trend observed for In (but not Zn) when comparing our CM volatility scale with that of Lodders (2003). Obvious causes for the observed variations between these scales cannot be extracted from observables at this time, however it is known the volatility of Zn changes as a function of pressure, being ~1500K at 1 bar and decreasing to ~850K at $10^{-5}$ bar (in silicates) (Schaefer and Fegley, 2010a), and that the volatility of elements is also heavily influenced by kinetic factors such as diffusion (Ikramuddin et al., 1977). As such, it is likely that a combination of factors (T, P, $fO_2$) has produced such differences between the relative volatility trends of CCs and the solar nebula. Additionally, the relative volatility scale(s) developed herein appear different to that determined by the experiments of Matza and Lipschutz (1977) (Table 4). As with the variations compared to solar nebula volatility, this suggests that the conditions of these experiments are different to those experienced by the heated CCs in this study, producing a re-arrangement of element volatility, notably for Zn in CM chondrites. A comparison of our volatility trend(s) to that of Matza and Lipschutz (1977) suggests that the heating conditions for the CMs PCA 02010 and PCA 02012 were at high temperature (at and above ~700°C), and therefore include the enhanced mobility of Zn observed at 700°C in their experimental study, an observation not visible when comparing condensation temperatures alone. When considering the volatile element depletion trend of Matza and Lipschutz (1977) for Murchison at 700°C, our data are in good agreement with theirs.

The relative volatility scales determined herein for the CM and CR chondrites are quite similar, with the exception that Zn and Tl have exchanged positions at opposite ends of the scale(s), i.e. Tl is the most volatile in the CMs and least volatile in the CRs, and the relative volatility of Zn decreases from CM to CR chondrites, and that the volatility of In is enhanced in the CR chondrites. If the relative volatilities of these elements are indeed different between the two chondrite groups, this would suggest that the CM and CR chondrites may have experienced thermal alteration under different conditions. A possible explanation for the change in volatility for In may lie in the fact that its volatility can change markedly as a function of $fO_2$, i.e. it displays a decrease in volatility with increasing $fO_2$ (O'Neill and Palme, 2008). This coupled with the observed high(er) water content of CM chondrites compared to CRs (e.g. Alexander et al., 2012; Garenne et al., 2014) may indicate that an ephemeral impact-generated atmosphere of a CM chondrite parent body would be more hydrous, and thus more oxidized, than that of a CR chondrite parent body (Schaefer and Fegley, 2010b; Garenne et al., 2014), thus decreasing the volatility of In in CM chondrites relative to CRs. Understanding the reversal of volatility for Zn and Tl is less straightforward. As previously discussed, the volatility of Zn is significantly enhanced at and above 700°C (Matza and Lipschutz, 1977) and at lower pressure (Schaefer and Fegley, 2010a), and therefore it's dampened volatility in the CR chondrites may indicate an overall lower temperature range and higher pressure for volatilization in the CR chondrites studied. Moreover, the volatility of elements such as Zn (and perhaps Bi and Sn), with a dominant oxidation state of 2+, decrease with increasing $fO_2$ and should be less volatile than Tl, which is monovalent in meteorites (Schaefer and Fegley, 2010a), in more oxidizing conditions (O'Neill and Palme, 2008), in qualitative agreement with our results for the CM chondrites (Table 4). It is also possible that this observation is due to some other control(s) - composition, thermal diffusion, the inclusion of GRO 95577. Most likely, many or all of these parameters impact the relative volatility of the elements, however the relative importance and efficacy of each parameter is beyond the scope of the current work. Determining the exact causality of these changes in volatility, and disentangling the inputs from these various parameters - as well as constraining other possible factors (e.g. composition, thermal diffusion, phyllosilicate fraction) - through further experimental work and discovery is at the forefront of future work.

The comparisons between our relative volatility trend(s) and those from thermodynamic considerations and experiments exemplify the power and utility of developing volatility scales from natural observations and extracting information from their similarities and differences. Developing such scales for the various meteorite types that are based on natural observations, and thus inherently account for the multitude of variables that may be (and likely are) at play during post-accretionary heating, such as $fO_2$, pressure, composition, diffusion, host phases, and temperature variability (Schaefer and Fegley, 2010a; Matza and Lipschutz, 1977), will further advance are understanding of the complex interplay of these controls during heating on meteorite parent bodies.

**4.2 Volatilization, dehydration, and open-system heating**

Volatile element abundances observed in the current study span a broad range, from canonical values to intensely depleted; moreover, water contents for samples in the current study range from case-normative to intensely depleted (Table 2). Fig. 4 displays the normalized abundances of volatile elements investigated in this study for CM chondrites as a function of water content ($H_2O$ wt%) as determined by Garenne et al. (2014), or calculated stoichiometrically from Alexander et al. (2012). Volatile element abundances are case-normative for the majority of the samples, and are congruent with literature data (Xiao and Lipschutz, 1992; Zolensky et al., 1992, 1997; Wang and Lipschutz, 1998). The constancy of volatile element contents across a range of water contents from ~5 wt% up to ~14 wt%, strongly reaffirms the notion that that the degree of aqueous alteration itself does not appear to affect bulk volatile element contents. Moreover, this constancy at water contents above ~5 wt%, which samples a spectrum of phyllosilicate to (oxy)hydroxide ratios (see Garenne et al., 2014) and weathering grades (see Table 1), further suggests that neither the type(s) of aqueous alteration nor the degree of terrestrial weathering significantly affects the volatile element contents. However, significant depletions in volatile elements are seen after a threshold water content of ~5 wt% in the current study (Fig. 4). The exact position of this threshold is unclear, however, as the meteorite EET 83355 (at 3.6 wt% $H_2O$), though sometimes reported as C2/C2M (Xia and Lipschutz, 1992; Alexander et al., 2013), is reported elsewhere (Wang and Lipschutz, 1998; Garenne et al., 2014) as an ungrouped C2, and therefore its initial elemental content may differ from the rest of the data set.

Many of the heated CM chondrites in Figure 4 display similar volatile element contents to the case-normative, unheated chondrites, and furthermore do not display enrichments in heavy isotopes of Zn, with $\delta^{66}$Zn values tightly clustered around case-normative values (Fig. 1 and Moynier et al., 2017). This indicates that the heating events these sample incurred were at viable temperatures for liberating H, C, and N, but not elements of lower volatility such as those investigated herein. This interpretation is consistent with canonical element volatility calculations (e.g. Lodders, 2003) for these elements, as well as with the heating classification scheme of Alexander et al. (2012/2013). That the heated ungrouped C2 chondrite EET 83355 (Nakamura, 2005; Alexander et al., 2012, 2013) and heated CV chondrite RBT 04133 (Davidson et al., 2014) do not display depletions in the elements studied herein and do not display enrichments in the heavy Zn isotopes is also consistent with this interpretation, as EET 83355 (EET 87522 as well) is estimated to have been heated below 300°C (Nakamura, 2005) and RBT 04133 to approximately 440°C (Davidson et al., 2014), temperatures below that necessary to volatilize Zn (or other elements studied herein) (e.g. Matza and Lipschutz, 1977). The curious data for the heated CR chondrite GRA 06100, with a nominally average Zn concentration and high Pb concentration, along with a negative $\delta^{66}$Zn, is suggestive of either heating above 700°C followed by disproportionate condensation of Zn and Pb back onto the parent body, or of large scale element mobilization due to impact heating, both of which could possibly produce these data. However, GRA 06100 requires further investigation in order to understand its complex history and substantiate any such hypotheses.

Of particular note are the two CM chondrites PCA 02010 and PCA 02012, as well as the CR chondrite GRO 03116, which all display highly depleted volatile element abundance patterns, and when we compare these results with water content data in literature (Garenne et al., 2014), we find that these samples are also significantly dehydrated. Importantly, the $\delta^{66}$Zn values for these samples determined in the current study are significantly enriched in the heavy Zn isotopes relative to normative values (Tables 1 and 2, Fig. 1), wherein enrichments greater than approximately 0.6 ‰ are difficult to explain by igneous processes (Moynier et al., 2017) and are indicative of evaporative processes (e.g. Moynier et al., 2009, 2011; Day et al., 2017a). This interpretation is also in line with that of Nakato et al. (2013), wherein a heating event at or above 900°C was inferred for PCA 02012. Taken together, all lines of evidence provide a strong argument for the occurrence of post-accretionary thermal alteration with enough energy to induce significant heating well above 700°C in these samples, thus liberating volatile elements such as Zn.

The three isotope plot in Fig. 5 displays $\delta^{68}$Zn and $\delta^{66}$Zn for all samples from Table 1, graphically illustrating that mass-dependent process(es) have produced all measured isotopic signatures in the current study, with a slope of 1.99 ($R^2$ = 0.99), in good agreement with theoretical predictions for a kinetic isotopic fractionation and with previous work (e.g. Luck et al., 2005; Chen et al., 2013; Pringle et al., 2017). Moreover, Fig. 5 clearly illustrates the exceptionally high $\delta^{66}$Zn values of the CM chondrites PCA 02010 and PCA 02012, and the CR chondrite GRO 03116, strongly indicating that these samples underwent volatile loss of Zn via a mass-dependent, kinetic process such as evaporation. Furthermore, there are no known samples with such isotopically heavy values as those measured in the samples discussed above (e.g. Luck et al., 2005; Pringle et al., 2017; Paniello et al., 2012ab; Herzog et al., 2009; Kato et al. 2015; Day et al., 2017b), and moreover the majority of terrestrial processes produce $\delta^{66}$Zn values less than 1 ‰ (Moynier et al., 2017), thus making terrestrial weathering an unlikely culprit for such values. Without any calculation, it is clear that the $\delta^{66}$Zn (and $\delta^{68}$Zn) values of PCA 02010 and PCA 02012 are incoherent with pure or non-ideal Rayleigh distillation - a possible evaporative regime - as a singular process, as this would require that PCA 02010 display the higher $\delta^{66}$Zn value of the two, and such is not the case (Fig. 1 and 5). The $\delta^{66}$Zn value obtained for PCA 02012 is in accord with a non-ideal kinetic isotope fractionation, wherein isotopic fractionation of Zn by evaporation is dampened due to inefficient and/or partial evaporation, leading to a higher fractionation factor (0.999, versus ~0.993 for Zn complexed to Cl or S in pure Rayleigh distillation) (see Day and Moynier (2014) for Zn isotopic fractionation via Rayleigh distillation in both regimes, and associated fractionation factors). A possible explanation for the limited isotopic fractionation of PC 02010 is that at the very low concentrations of PCA 02010, even a minute amount of isotopically lighter Zn contamination could have severely dampened its heavy isotopic signature. Another viable possibility is that PCA 02010 incurred Zn evaporation at a higher pressure than PCA 02012 in a kinetic mass-dependent fractionation regime, which would likely cause a faster rate of evaporation concomitant with attenuated enrichment in the heavy isotopes of Zn (Young et al., 1998). Regardless of evaporative regime or any contamination that may have dampened, or otherwise affected, the high $\delta^{66}$Zn value of PCA 02010, intense isotopic fractionation of Zn via evaporation remains the most plausible explanation for the $\delta^{66}$Zn values observed for PCA 02010, as well as for the CM chondrite PCA 02012 and the CR chondrite GRO 03116. The heavy Zn isotopic signatures of these samples provide very robust evidence that whatever thermal event(s) these samples incurred happened in an open system, thus allowing for the intense isotopic fractionation of Zn observed in this study.

**4.3 Sources and conditions for heating and volatile transport in CM and CR chondrites**

The dynamics of parent body accretion are generally thought to be low-energy, "gentle" impacts, and therefore accretionary processes are unlikely to be a source of significant heating in chondrite parent bodies (Melosh, 1990; Wood and Pellas, 1991; Ghosh et al., 2006). Solar winds are focused at the poles of T Tauri stars and not the matter-inhabited mid-plane, thus it is also unlikely that electromagnetic induction could have supplied enough energy to significantly heat these meteorite parent bodies (Wood and Pellas, 1991; Ghosh et al., 2006; Huss et al., 2006). Radioactive heating by short-lived radionuclides – namely $^{26}$Al and $^{60}$Fe – cannot be disregarded as a possible heat source, however it is difficult to reconcile the intense volatile element depletions and heavy Zn isotopic signatures observed in the current study with the canonical onion shell model for thermal metamorphism (e.g. Trieloff et al., 2003), wherein internal heating via radioactive decay occurs and metamorphic grade decreases radially from the center of the parent body. Moreover, radioactive heating, for example by the decay of $^{26}$Al, has a minimum requisite heating time of 0.7 million years (one half-life), a duration which engenders Fe-Mg inter-diffusion between chondrules and matrix (Nakamura et al, 2006), a feature not exhibited in the CM chondrite PCA 02012 (Nakato et al., 2013) or the CR chondrite GRO 03116 (Schrader et al., 2015). Solar radiative heating via an orbit passing inside ~ 0.1 AU is another potential heat source for solar system materials, is similar to radioactive decay in duration and requisite peak temperatures for dehydrating phyllosilicates, and thus bares similar mineralogical and chemical consequences, however its timing is different, being later than radioactive heating (Nakamura, 2005). Again, the lack of evidence PCA 02012 and GRO 03116 for long duration heating render this an unlikely heat source. Therefore, none of the preceding potential sources seem to produce viable conditions on short enough timescales to produce the cumulative observations for the most water and volatile depleted samples within our study.

Shock metamorphism via impact, on the other hand, has been invoked to explain the low water contents and peculiar coloration of PCA 02010 and PCA 02012 (Garenne et al., 2014), the maturation of organics and overall petrography of PCA 02012 (Nakato et al., 2013) and indicated via petrographic evidence in GRO 03116 (melt veining, compacted matrix and foliation) (Abreu, 2011; Schrader et al., 2015). Therefore, shock metamorphism is the most plausible and most likely energy source (e.g. Bischoff and Stöffler, 1992; Nakamura, 2005) for these carbonaceous chondrites, as it is capable of producing both the volatile element depletions and the large isotopic fractionations in Zn (e.g. Moynier et al., 2010, 2017) observed in these samples without producing Fe-Mg diffusion characteristic of long(er) duration heating (e.g. Nakamura, 2006; Schrader et al., 2015), such as that seen in thermally altered ordinary chondrites (Nakamura et al., 2006). Furthermore, impact shock as a process is capable of producing peak temperatures well in excess of 700°C (beyond 2000°C in porous materials such as CM chondrites), and can thusly liberate volatile elements to the extent exhibited in the most water and volatile depleted samples of the current study (Bischoff and Stöffler, 1992; Beck et al., 2007; Garenne et al., 2014). Additionally, heating by impact is a relatively short-duration, localized process, i.e. one that affects only the body that incurs it (possibly not even in its entirety) and is thus non-ubiquitous in the solar system, a fact in line with the exceptionality of heated CM and CR chondrites. Therefore, volatile loss by evaporation due to impact heating on a CM or CR parent body stands out as the most plausible scenario, an interpretation that is also consistent with the recent observation of Si isotopic fractionation in angrite meteorites, which show variables enrichments in the heavier isotopes of Si (Pringle et al., 2014). Moreover, it has been shown that impact shocks were at the origin of volatile element variability and isotopic composition of Zn in ureilites, in which $\delta^{66}$Zn strongly correlates with the shock grade, and the highly shocked samples displayed the heaviest $\delta^{66}$Zn values concomitant with the lowest Zn abundances (Moynier et al., 2010). Terrestrial analogs, such as tektites and trinitites (Moynier et al., 2009; Day et al., 2017a, respectively) mirror such observations, wherein the volatile-depleted glasses produced show significant enrichments in the heavy isotopes of Zn. Further to this point, such enrichments in heavy Zn isotopes are generally not seen in ordinary chondrites, which often dominantly display $\delta^{66}$Zn values less than 1 ‰ (Moynier et al., 2017), suggesting a fundamental difference in the prevalent source of heat. The observed strong enrichment in heavy Zn isotopes in the intensely volatile depleted CM chondrites PCA 02010 and PCA 02012, as well as the CR chondrite GRO 03116, are in accord with such observations, and their linkage to impact shock processes provide robust evidence that shock heating was the prevalent source of their volatile depletion and intense Zn isotopic fractionation. In summary, the observations of the current study provide very strong evidence that post-accretionary impact shock was the dominant heat source for these carbonaceous chondrites. These observations in conjunction with insights from other solar system materials and terrestrial analogs suggest that, contrary to the ordinary chondrites, which suffered appreciably longer (and earlier) heating events (Nakamura et al., 2006), the most likely and prevalent heat source for thermal alteration on CCs is post-accretionary impact shock.

## 5. Conclusions

We have reported MC-ICPMS Zn isotopic data and HR-ICPMS trace element data for a suite of CM and CR carbonaceous chondrites. We have used trace element data for Zn, In, Sn, Tl, Pb and Bi in conjunction with literature data in order to develop and apply an empirical method for determining the relative volatility of these elements during post-accretionary heating in the CM and CR chondrites. These results indicate that the order of volatility in CMs is Pb-Sn-Bi-In-Zn-Tl, and in CRs is Tl-Zn-Sn-Pb-Bi-In. This order is different from the scale obtained under nebular condition (Bi-Pb-Zn-Sn-In-Tl) and suggest that thermal alteration in the CM (PCA 02010, PCA 02012) and CR (GRO 03116) chondrites occurred under different conditions than that of their formation. Furthermore, the difference in the sequence of depletion between CM and CR chondrites suggests that they may have experienced thermally driven element mobilization and dehydration under different conditions, and specifically that the CM chondrites may have been heated under more oxidative conditions than the CR chondrites. A comparison of our elemental data with water content data from literature (Alexander et al., 2012, 2013; Garenne et al., 2014) shows a distinct relationship between volatile element concentrations in the CM chondrites (and to a lesser constrained degree in the CRs) and their water contents. This observation provides very strong evidence that a direct link exists between the volatile elements and water contents of carbonaceous chondrites, and suggests that the same may then be true for larger planetary bodies such as Earth. Moreover, while previous studies (e.g. Paul and Lipschutz, 1989; Garenne et al., 2014) tentatively invoked open-system thermal alteration as the most straightforward means to account for volatile element and water depletions in carbonaceous chondrites, our study provides concrete evidence that these thermal events took place in an open system, as it was found that the most volatile and water depleted samples are also those displaying remarkably high $\delta^{66}$Zn values, suggesting preferential loss of light Zn isotopes to the gas phase during open-system evaporation. The Zn isotopic data from this study, linked with water content and petrologic data, strongly indicate that impact shock was the most likely and dominant heat source for the CM and CR chondrites. Moreover, these observations highlight the necessity of – and opportunity for – more observational and experimental work aimed at investigating the volatility of elements in meteoritic materials under more relevant conditions, i.e. under higher pressures and oxygen fugacities (relative to the solar nebula) that most likely prevailed during heating events on carbonaceous chondrite parent bodies and other telluric bodies, as well as incorporating kinetic processes such as diffusion through varying compositional and spatial contexts. As the origin and eventual fate of these elements in meteorites and Earth are intimately linked, progressing our understanding of their volatility during post-accretionary thermal processing in meteorites will provide us with more direct links to these processes on Earth and other planetary bodies.

**Acknowledgments**

The authors very kindly thank Dr. Devin Schrader and two anonymous reviewers, whose input has substantially enhanced the framework and discussion of the manuscript. We also sincerely thank James Day, Manuel Moreira, Bruce Fegley, Katharina Lodders, Paolo Sossi, and Zhengbin Deng for fruitful discussions. BM thanks Julien Moureau, and Mickaël Tharaud for their respective expertise with MC-ICPMS and Element 2 mass spectrometry, as well as Delphine Limmois and Laeticia Faure for their help in the clean laboratory. BM and EP thank the financial support of the IDEX USPC through PhD fellowships. JS and BM thank the financial support of the French National Research Agency (ANR Project VolTerre, grant no. ANR-14-CE33-0017-01). FM and JS thank the financial support of the UnivEarthS Labex program at Sorbonne Paris Cité (ANR-10-LABX-0023 and ANR-11-IDEX-0005-02). FM acknowledges funding from the European Research Council under the H2020 framework program/ERC grant agreement #637503 (Pristine), and the ANR through a chaire d'excellence Sorbonne Paris Cité. Parts of this work were supported by IPGP multidisciplinary program PARI, and by Region Île-de-France SESAME Grant no. 12015908.

**References**

Abreu N.M. (2011) Petrographic evidence of shock metamorphism in CR2 chondrite GRO 03116 (abstract #5211). *Meteoritics & Planetary Science 46:A4*.

Abreu N.M. and Bullock E.S. (2013) Opaque assemblages in CR2 Graves Nunataks (GRA) 06100 as indicators of shock-driven hydrothermal alteration in the CR chondrite parent body. *Meteoritics and Planetary Science* **48**, 2406-2429.

Akai J. (1992) T-T-T diagram of serpentine and saponite, and estimation of metamorphic heating degree of Antarctic carbonaceous chondrites. *Antarct. Meteor. Res.* **5**, 120-135.

Alexander C.M. O'D., Bowden R., Fogel M.L., Howard K.T., Herd C.D.K, Nittler L.R. (2012) The provenances of asteroids, and their contributions to the volatile inventories of the terrestrial planets. *Science* **337**, 721-723.

Alexander C.M. O'D., Howard K.T., Bowden R., Fogel M.L. (2013) The classification of CM and CR chondrites using bulk H, C, and N abundances and isotopic compositions. *Geochim. Cosmochim. Acta* **123**, 244-260.

Anders E. (1964) Origin, age, and composition of meteorites. *Space Sci. Rev.* **3**, 583-714.

Barrat J. A., Zanda B., Moynier F., Bollinger C., Liorzou C., Bayon G. (2012) Geochemistry of CI chondrites: Major and trace elements, and Cu and Zn Isotopes. *Geochim. Cosmochim. Acta* **83**, 79–92.

Beck P., De Andrade V., Orthous-Daunay F-R., Veronesi G., Cotte M., Quirico E., Schmitt B. (2012) The redox state of iron in the matrix of CI, CM, and metamorphosed CM chondrites by XANES spectroscopy. *Geochim. Cosmochim. Acta* **99**, 305-316.

Beck P., Ferroir T., Gillet P. (2007) Shock-induced compaction, melting, and entrapment of atmospheric gases in Martian meteorites. *Geophys. Res. Lett.* **34**, L01203.

Beck P., Garenne A., Quirico E., Bonal L., Montes-Hernandez G., Moynier F., Schmitt B. (2014) Transmission infrared spectra (2-25μm) of carbonaceous chondrites (CI, CM, CV-CK, CR, C2 ungrouped): mineralogy, water, and asteroidal processes. *Icarus* **229**, 263-277.

Bischoff A. (1998) Aqueous alteration of carbonaceous chondrites: evidence for preaccretionary alteration – a review. *Meteoritics and Planetary Science* **33**, 1113-1122.

Bischoff A. and Stöffler D. (1992) Shock metamorphism as a fundamental process in the evolution of planetary bodies: information from meteorites. *Eur. J. Mineral.* **4**, 707-755.

Bland P.A., Alard O., Benedix G.K., Kearsley A.T., Menzies O.N., Watt L.E., Rogers N.W. (2005) Volatile fractionation in the early solar system and chondrule/matrix complementarity. *Proceedings of the National Academy of Sciences* **102** (39), 13755-13760.

Brearley A.J. (2006) The action of water. In *Meteorites and the Early Solar System II* (ed. Lauretta D.S., McSween (Jr.) H.Y). University of Arizona Press, Tucson, pp. 587-624.

Briani G., Quirico E., Gounelle M., Paulhiac-Pison M., Montagnac G., Beck P., Orthous-Daunay F-R., Bonal L., Jacquet E., Kearsley A., Russell S.S. (2013) Short duration thermal metamorphism in CR chondrites. *Geochim. Cosmochim. Acta* **122**, 267–279.

Chen H., Nguyen B.M., Moynier F. (2013a) Zinc isotopic composition of iron meteorites: absence of isotopic anomalies and origin of the volatile element depletion. *Meteoritics and Planetary Science* **48**, 2441-2450.

Choe W.H., Huber H., Rubin A.E., Kallemeyn G.W., Wasson J.T. (2010) Compositions and taxonomy of 15 unusual carbonaceous chondrites. *Meteoritics and Planetary Science* **45**, 531-554.

Clayton R.N. and Mayeda T.K. (1999) Oxygen isotope studies of carbonaceous chondrites. *Geochim. Cosmochim. Acta* **63**, 2089–2104.

Cloutis E.A., Hudon P., Hiroi T., Gaffey M.J., Mann P. (2011) Spectral reflectance properties of carbonaceous chondrites 2: CM chondrites. *Icarus* **216**, 309-346.

Cloutis E.A., Hudon P., Hiroi T., Gaffey M.J. (2012) Spectral reflectance properties of carbonaceous chondrites 4: aqueously altered and thermally metamorphosed meteorites. *Icarus* **220**, 586-617.

Davidson J., Schrader D.L., Busemann H., Franchi I.A., Connolly H.C. Jr., Lauretta D.S., Alexander C.M.O.'D., Verchovsky A., Greenwood R.C. (2014) Petrography, stable isotope compositions, microRaman spectroscopy and presolar components of Roberts Massif 04133: a reduced CV3 carbonaceous chondrite. *Meteoritics and Planetary Science* **49**, 2133-2151.

Day J.M.D. and Moynier F. (2014) Evaporative fractionation of volatile stable isotopes and their bearing on the origin of the moon. *Philos. Trans. R. Soc. A* **372**, 20130259.

Day J.M.D., Moynier F., Meshik A.P., Pradivtseva O.V., Petit D.R. (2017a) Evaporation fractionation of zinc during the first nuclear detonation. *Science Advances* **3**, e1602668.

Day J.M.D., Moynier F., Shearer C.K. (2017b) Late-stage magmatic outgassing from a volatile-depleted Moon. *Proceedings of the National Academy of Sciences* **114** (36), DOI: 10.1073.

Gale A., Dalton C.A., Langmuir C.H., Su Y., Schilling J-G. (2013) The mean composition of ocean ridge basalts. *Geochemistry, Geophysics, Geosystems* **14**, 489-518.

Garenne A., Beck P., Montes-Hernandez G., Chiriac R., Toche F., Quirico E., Bonal L., Schmitt B. (2014) The abundance and stability of "water" in type 1 and 2 carbonaceous chondrites (CI, CM, and CR). *Geochim. Cosmochim. Acta* **137**, 93-112.

Ghosh A., Weidenschilling S.J., McSween H.Y. Jr., Rubin A. (2006) Asteroidal heating and thermal stratification of the asteroid belt. In *Meteorites and the Early Solar System II* (ed. Lauretta D.S., McSween (Jr.) H.Y). University of Arizona Press, Tucson, pp. 555-566.

Herzog G.F., Moynier F., Albarède F., Berezhnoy A.A. (2009) Isotopic and elemental abundances of copper and zinc in lunar samples, Pele's hairs, and a terrestrial basalt. *Geochim. Cosmochim. Acta* **73**, 5884-5904.

Howard K.T., Alexander C.M.O.'D., Schrader D.L., Dyl K.A. (2015) Classification of hydrous meteorites (CR, CM, and C2 ungrouped) by phyllosilicate fraction: PSD-XRD modal mineralogy and planetesimal enironments. *Geochim. Cosmochim. Acta* **149**, 206–222.

Huss G.R., Meshik A.P., Smith J.B., Hohenberg C.M. (2003) Presolar diamond, silicon carbide, and graphite in carbonaceous chondrites: Implications for thermal processing in the solar nebula. *Geochim. Cosmochim. Acta* **67**, 4823-4848.

Huss G.R., Rubin A.E., Grossman J.N. (2006) Thermal metamorphism in chondrites. In *Meteorites and the Early Solar System* (ed. Lauretta D.S., McSween (Jr.) H.Y). University of Arizona Press, Tucson, pp. 567-586.

Ikramuddin M., Matza S., Lipschutz M.E. (1977) Thermal metamorphism of primitive meteorites – V. Ten trace elements in Tieschitz H3 chondrite heated at 400-1000°C. *Geochim. Cosmochim. Acta* **41**, 1247-1256.

Kato C., Moynier F., Valdes M.C., Dhaliwal J.K., Day J.M.D. (2015) Extensive volatile loss during formation and differentiation of the Moon. *Nat. Comm.* **6** (7617). DOI: 10.1038/ncomms8617.

Lange M.A. and Ahrens T.J. (1982) The evolution of an impact-generated atmosphere. *Icarus* **51**, 96-120.

Lodders K. (2003) Solar system abundances and condensation temperatures of the elements. *Astrophys. J.* **591**, 1220–1247.

Lodders K. and Fegley (Jr.), B. (1998) In *The Planetary Scientist's Companion*. Oxford Univ. Press, Oxford. pp. 311-317.

Luck J.M., Ben Othman D., Albarède F. (2005) Zn and Cu isotopic variations in chondrites and iron meteorites: early solar nebula reservoirs and parent-body processes. *Geochim. Cosmochim. Acta* **69**, 5351-5363.

Mahan B., Siebert J., Pringle E.A., Moynier F. (2017) Elemental partitioning and isotopic fractionation of Zn between metal and silicate and geochemical estimation of the S content of the Earth's core. *Geochim. Cosmochim. Acta* **196**, 252-270.

Marechal N., Telouk P., Albarède F. (1999) Precise analysis of copper and zinc isotopic compositions by plasma-source mass spectrometry. *Chem. Geol.* **156**, 251–273.

Matza S.D., Lipschutz M.E. (1977) Thermal metamorphism of primitive meteorites – VI. Eleven trace elements in Murchison C2 chondrite heat at 400-1000°C. *Proc. Lunar Sci. Conf.* **8**, 161-176.

Melosh, H.J. (1990) Giant impacts and the thermal state of the early Earth. In *Origin of the Earth* (ed. H.E. Newsom and J.H. Jones), Oxford University, New York, pp. 69-84.

Moynier F., Albarède F., Herzog G.F. (2006) Isotopic composition of zinc, copper, and iron in lunar samples. *Geochim. Cosmochim. Acta* **70**, 6103–6117.

Moynier F. and Le Borgne M. (2015) High precision zinc isotopic measurements applied to mouse organs. *J. Vis. Exp.* **99**, e52479.

Moynier F., Beck P., Jourdan F., Yin Q.-Z., Reimold U., Koeberl C. (2009) Isotopic fractionatino of Zn in tektites. *Earth Planet. Sci. Lett.* **277**, 482–489.

Moynier F., Beck P., Yin Q.-Z., Ferroir T., Barrat J.-A., Paniello R., Telouk P., Gillet P. (2010) Volatilization induced impacts recorded in Zn isotope composition of ureilites. *Chem. Geol.* **276**, 374–379.

Moynier F., Paniello R.C., Gounelle M., Albarède F., Beck P., Podosek F., Zanda B. (2011) Nature of volatile depletion and genetic relationships in enstatite chondrite and aubrites inferred from Zn isotopes. *Geochim. Cosmochim. Acta* **75**, 297-307.

Moynier F., Vance D., Fujii T., Savage P. (2017) The isotope geochemistry of Zinc and Copper. *Rev. Mineral. Geochem.* **82**, 543-600.

Nakamura T. (2005) Post-hydration thermal metamorphism of carbonaceous chondrites. *Journal of Min. and Petr. Sci.* **100**, 260-272.

Nakamura T. (2006a) Yamato 793321 CM chondrite: dehydrated regolith material of a hydrous asteroid. *Earth Planet. Sci. Lett.* **242**, 26–38.

Nakamura T., R. Okazaki, Huss G.R. (2006b) Thermal metamorphism of CM carbanaceous chondrites: effects on phyllosilicate mineralogy and presolar grain abundances. *Lunar Planet. Sci. Conf.* (A#1633).

Nakato A., Brearley A.J., Nakamura T., Noguchi T., Ahn I., Lee J.I., Matsuoka M., Sasaki S. (2013) PCA 02012: a unique thermally metamorphosed carbonaceous chondrite. *Lunar Planet. Sci.* **44**, A2708.

O'Neill H.S.C. and Palme H. (1998) Composition of the silicate Earth: implications for accretion and core formation. In *The Earth's Mantle: Structure, Composition and Evolution – The Ringwood Volume* (ed. I. Jackson), Cambridge University Press, Cambridge.

O'Neill H.S.C. and Palme H. (2008) Collisional erosion and the non-chondritic composition of the terrestrial planets. *Philos. Trans. R. Soc. A* **366**, 4205-4238.

Palme H. and O'Neill H.S.C. (2014) Cosmochemical estimates of mantle composition. Chapter 3.1 in *Treatise on Geochemistry, second edition* (ed. K.K. Turekian), Elsevier, Oxford, UK.

Paniello R., Day J.M.D., Moynier F. (2012a) Zinc isotopic evidence for the origin of the Moon. *Nature* **490**, 376-380.

Paniello R., Moynier F., Beck P., Barrat J-A., Podosek F.A., Pichat S. (2012b) Zinc isotopes in HEDs: clues to the format of 4-Vesta, and the unique composition of Pecora Escarpment 82502. *Geochim. Cosmochim. Acta* **86**, 76-87.

Paul R.L., Lipschutz M.E. (1989) Labile trace elements in some antarctic carbonaceous chondrites: antarctic and non-antarctic meteorite comparisons. *Z Naturforsch* **44a**, 979-987.

Pringle E.A., Moynier F., Savage P.S., Badro J., Barrat J.-A. (2014) Silicon isotopes in angrites and volatile loss in planetesimals. *Proceedings of the National Academy of Sciences* **111** (48), DOI: 10.1073.

Pringle E.A., Moynier F. (2017) Rubidium isotopic composition of the Earth, meteorite, and the Moon: Evidence for the origin of volatile loss during planetary accretion. *Earth Planet. Sci. Lett.* **473**, 62–70.

Rubin A.E., Trigo-Rodrígruez J.M., Huber H., Wasson J.T. (2007) Progressive aqueous alteration of CM carbonaceous chondrites. *Geochim. Cosmochim. Acta* **71**, 2361-2382.

Schaefer L. and Fegley Jr. B. (2010a) Volatile element chemistry during metamorphism of ordinary chondritic material and some of its implications for the composition of asteroids. *Icarus* **205**, 483-496.

Schaefer L. and Fegley Jr. B. (2010b) Chemistry of atmsospheres formed during accretion and the Earth and other terrestrial planets. *Icarus* **208**, 438-448.

Schrader D.L., Connolly H.C., Lauretta D.S., Zega D.L., Davidson J., Domanik K.J. (2015) The formation and alteration of the Renazzo-like carbonaceous chondrites III: toward understanding the genesis of ferromagnesian chondrules. *Meteoritics and Planetary Sciences* **50**, 15-50.

Schrader D. L., Franchi I. A., Connolly H. C. Jr., Greenwood R. C., Lauretta D. S. and Gibson J. M. (2011) The formation and alteration of the Renazzo-like carbonaceous chondrites I: implications of bulk-oxygen isotopic composition. *Geochim. Cosmochim. Acta* **75**, 308-325.

Shu F. H., Shang H., Lee T. *(1996)* Towards an astrophysical theory of chondrites. *Science 271, 1545–1552.*

Tharaud M., Gardoll S., Khelifi O., Benedetti M.F., Sivry Y. (2015) uFREASI: user-Friendly Elemental dAta procesSIing. A free and easy-to-use tool for elemental data treatment. *Microchemical Journal* **121**, 32-40.

Tonui E.K., Zolensky M.E., Lipschutz M.E., Wang M.-S., Nakamura T. (2003) Yamato 86029: aqueously altered and thermally metamorphosed CI-like chondrite with unusual texture. *Meteoritics and Planetary Sciences* **38**, 269-292.

Tonui E.K., Zolensky M.E., Hiroi T., Nakamura T., Lipschutz M.E., Wang M.-S., Okudaira K. (2014) Petrographic, chemical and spectroscopic evidence for thermal metamorphism in carbonaceous chondrites I: CI and CM chondrites. *Geochim. Cosmochim. Acta* **126**, 284-306.

Trieloff M., Jessberger E.K., Herrwerth I., Hopp J., Fiéni C., Ghélis M., Bourot-Denise M., Pellas M. (2003) Structure and thermal history of the H-chondrite parent asteroid revealed by thermochronometry. *Nature* **422**, 502-506.

Van Schmus W.R. and Wood J.A. (1967) A chemical-petrologic classification for the chondritic meteorites. *Geochim. Cosmochim. Acta* **31**, 747–765.

Wang M.-S. and Lipschutz M.E. (1998) Thermally metamorphosed carbonaceous chondrites from data for thermally mobile trace elements. *Meteoritics and Planetary Sciences* **33**, 1297-1302.

Wasson J.T. and Chou C-L. (1974) Fractionation of moderately volatile elements in ordinary chondrites. *Meteoritics* **9**, 69-84.

Weisberg M.K. and Huber H. (2007) The GRO 95577 CR1 chondrite and hydration of the CR parent body. *Meteoritics and Planetary Sciences* **42**, 1495-1503.

Weisberg M.K., McCoy T.J., Krot A.N. (2006) Systematics and evaluation of meteorite classification. In *Meteorites and the Early Solar System II* (ed. Lauretta D.S., McSween (Jr.) H.Y). University of Arizona Press, Tucson, pp. 19-52.

Weisberg M.K., Smith C., Benedix G., Folco L., Righter K., Zipfel J., Yamaguchi A., Chennaoui Aoudjehane H. (2008) The meteoritical bulletin, no. 94, September 2008. *Meteoritics and Planetary Sciences* **43**, 1551-1588.

Wood J.A. and Pellas P. (1991) What heated the meteorite planets? In *The Sun in Time* (ed. Sonett C.P. et al.) University of Arizona Press, Tucson, pp. 740-760.

Xiao X. and Lipschutz M.E. (1992) Labile trace elements in carbonaceous chondrites: a survey. *J. Geophys. Res.* **97**, 10199-10211.

Yin Q-Z. (2005) From dust to planets: The tale told by moderately volatile elements. In *Chondrites and the Protoplanetary Disk*, ASP Conference Series (eds. Krot A.N. et al.) ASP Conference Series **341**, 632-644.

Young E., Nagahara H., Mysen B.O., Audet D.M. (1998) Non-Rayleigh oxygen isotope fractionation by mineral evaporation: Theory and experiments in the system SiO2. *Geochim. Cosmochim. Acta* **62**, 3109-3116.

Zolensky M.E., Hewins R.H., Mittlefehldt D.W., Lindstrom M. M., Xiao X., Lipschutz M. E. (1992) Mineralogy, petrology, and geochemistry of carbonaceous chondrite clasts in the LEW 85300 polymict eucrite. *Meteoritics* **27**, 596-604.

Zolensky M.E., Mittlefehldt D.W., Lipschutz M.E., Wang M-S., Clayton R.N., Mayeda T.K., Grady M.M., Pillinger C., Barber D. (1997) CM chondrites exhibit the complete petrologic range from type 2 to 1. *Geochim. Cosmochim. Acta* **61**, 5099-5115.

**Figure Captions**

Figure 1: $\delta^{66}$Zn as a function of water content for CM chondrites (water content from Garenne et al., 2014). Errors for all samples are two times the standard deviation (2σ). Solid black line indicates average CM $\delta^{66}$Zn value (~0.38‰, Moynier et al., 2017). Scale for $\delta^{66}$Zn is logarithmic to ease viewing (note that this affects relative size of error bars). Zinc isotopes display nominally average values down to 3.6 wt% $H_2O$, below which samples PCA 02010 and PCA 02012 display a $\delta^{66}$Zn signal heavily enriched in the heavy isotopes, suggesting evaporative processes and preferential loss of isotopically light Zn to the gas phase.

Figure 2: (a-e) M/Zn plots for volatile elements in CM chondrites. Data in open and black symbols are from this study; data in grey are from literature (Paul and Lipschutz, 1989; Xiao and Lipschutz, 1992; Zolensky et al., 1992, 1997; Wang and Lipschutz, 1998). Normalized concentrations on the y-axis are denoted by the element symbol at top left; normalized Zn concentrations are on the x-axis for all plots. All data have been normalized using local maxima (see *Section 4.1*). The largest slope (Tl) indicates that the element is the most volatile compared to Zn (and vice versa), and a slope of unity is the relative volatility of Zn. Note that for Pb and Sn, only data from this study were available and thus input into calculation. Figure 2f displays the slopes for elements in CM chondrites as a function of their y-intercepts.

Figure 3: (a-e) M/Zn plots for volatile elements in CR chondrites. Data in open and black symbols are from this study. Normalized concentrations on the y-axis are denoted by the element symbol at top left; normalized Zn concentration is on the x-axis for all plots. All data have been normalized using local maxima (see *Section 4.1*). The largest slope (In) indicates that the element is the most volatile compared to Zn (and vice versa), and a slope of unity is the relative volatility of Zn. Note that the CR1 GRO 95577 contains the highest concentrations for all elements investigated, and thus has a strong influence on determination of the relative volatility of the elements. Figure 3f displays the slopes for elements in CR chondrites as a function of their y-intercepts.

Figure 4: Normalized abundance data for volatile elements in the CM chondrites as a function of water content (Garenne et al., 2014). Data in open and black symbols are from this study; data in grey are from literature - ALH 81002 data from Xia and Lipschutz (1992); Murchison data from Zolensky et al. (1992); Y 791198 data from Wang and Lipschutz (1998). Elements are ordered in the legend from least volatile to most, with $T_{50}$ from Lodders (2003) in parentheses for comparison. The volatile element abundances display nominally average values until a threshold content of around 5.5 wt% $H_2O$, below which significant depletions in all elements occur. This severe depletion is interpreted as loss of the elements via volatilization. The high water contents and relatively low volatile element abundance of ALH 83100 is possibly due to its abnormally high carbonate abundance (Zolensky et al., 1997), which may be depleted in volatiles relative to the phyllosilicates.

Figure 5: Three isotope plot of $\delta^{68}Zn$ versus $\delta^{66}Zn$ for all samples analyzed in the current study. Diamonds are CM chondrites; circles are CR chondrites. Errors for all samples are two times the standard deviation ($2\sigma$); due to scaling, errors for PCA 02010, PCA 02012 and GRO 03116 are within the symbol. For visibility, inlayed subplot displays data with case-normative $\delta^{68}Zn$ versus $\delta^{66}Zn$ values, which are all effectively contained at the bottom left corner. Dashed black line indicates the calculated slope of mass-dependent isotopic fractionation (m = 1.99). The CM chondrites PCA 02010 PCA 02012, and the CR chondrite GRO 03116, are intensely enriched in the heavy isotopes of Zn and are thus distinctly separated from the other samples along the line of mass-dependent fractionation.

Table 1 – Water content and Zn isotopic signature of CM-like (above partition) and CR/CV (below partition) chondrites by MC-ICPMS

| Sample[a] | Type[b] | Weathering[b] | H$_2$O (wt%)[c] | δ$^{66}$Zn | 2σ | δ$^{68}$Zn | 2σ | n |
|---|---|---|---|---|---|---|---|---|
| ALH 83100 | CM1/2 | Be | 13.9 | 0.32 | 0.10 | 0.71 | 0.18 | 3 |
| *ALH 84033* | CM2 | Ae | 6.9 | 0.43 | 0.04 | 0.80 | 0.10 | 3 |
| ALH 84044 | CM2 | Ae | 12.9 | 0.32 | 0.02 | 0.59 | 0.05 | 3 |
| *DOM 03183* | CM2 | B | 9.6 | 0.32 | 0.08 | 0.65 | 0.12 | 3 |
| *EET 87522* | CM2 | Be | 5.5 | 0.33 | 0.03 | 0.65 | 0.06 | 3 |
| *EET 96029* | CM2 | A/B | 8.2 | 0.33 | 0.07 | 0.65 | 0.12 | 4 |
| LAP 03718 | CM2 | BE | 9.8 | 0.32 | 0.03 | 0.62 | 0.03 | 3 |
| LEW 85311 | CM2 | Be | 9.9 | 0.32 | 0.03 | 0.60 | 0.12 | 3 |
| LEW 85312 | CM2 | B | 8.2 | 0.35 | 0.06 | 0.69 | 0.10 | 4 |
| LEW 87022 | CM2 | B | 12.4 | 0.30 | 0.01 | 0.66 | 0.06 | 3 |
| LEW 90500 | CM2 | B | 10.4 | 0.32 | 0.10 | 0.61 | 0.15 | 3 |
| LON 94101 | CM2 | Be | 10.6 | 0.32 | 0.01 | 0.63 | 0.04 | 3 |
| *MAC 88100* | CM2 | Be | 11.8 | 0.31 | 0.05 | 0.60 | 0.08 | 3 |
| *MIL 05152*[d] | CM2 | B | 5.50 | 0.45 | 0.07 | 0.83 | 0.17 | 3 |
| MCY 05230 | CM2 | B | 11.0 | 0.34 | 0.07 | 0.65 | 0.09 | 4 |
| MET 01070 | CM1 | Be | 11.1 | 0.30 | 0.07 | 0.59 | 0.12 | 3 |
| *MIL 07700* | CM2 | A | 5.5 | 0.31 | 0.07 | 0.59 | 0.06 | 3 |
| PCA 02010 | CM2 | B | 1.1 | 5.42 | 0.44 | 10.69 | 1.30 | 2 |
| PCA 02012 | CM2 | B | 1.7 | 16.73 | 0.09 | 33.30 | 0.20 | 4 |
| QUE 97990 | CM2 | BE | 8.7 | 0.34 | 0.04 | 0.70 | 0.10 | 4 |
| *EET 83355* | C2-ung | A/B | 3.6 | 0.36 | 0.06 | 0.77 | 0.13 | 3 |
| BUC 10933 | CR2 | B/Ce | - - - | 0.26 | 0.06 | 0.46 | 0.11 | 3 |
| *GRA 06100* | CR2 | B | 0.0 | -1.02 | 0.08 | -2.14 | 0.13 | 3 |
| *GRO 03116* | CR2 | B/C | 1.2 | 4.77 | 0.03 | 9.30 | 0.09 | 3 |
| GRO 95577 | CR1 | B | 12.3 | 0.33 | 0.02 | 0.59 | 0.04 | 3 |
| *LAP 02342*[d] | CR2 | A/B | 3.1 | 0.09 | 0.01 | 0.12 | 0.01 | 2 |
| MET 00426 | CR2 | B | 1.4 | 0.13 | 0.04 | 0.16 | 0.06 | 3 |
| *RBT 04133* | CV3 | B/C | 1.5 | 0.11 | 0.04 | 0.19 | 0.08 | 3 |

[a] Samples in *italics* are those identified as heated by: Alexander et al. (2012) for CMs, Abreu (2011) for GRA 06100 and GRO 03116, Davidson et al. (2014) for RBT 04133.
[b] Petrologic type and weather grade from Meteoritical Bulletin (www.lpi.usra.edu/meteor/metbull.php).
[c] Water loss in the 200-770ºC range determined by Garenne et al. (2014). This temperature range was chosen as it represents the dehydration of oxy-hydroxide minerals and hydroxyl groups from phyllosilicates, and excludes adsorbed water as well as de-carbonization and sulfate release.
[d] Calculated stoichiometrically from H contents (Alexander et al., 2012) and a 1/8 mass ratio of H to O in water.

Table 2 – Water content, δ$^{66}$Zn, and HR-ICPMS data for CM-like (above partition) and CR/CV (below partition) chondrites

| Sample[a] | H2O (wt%)[b] | δ$^{66}$Zn | 2σ | n | Zn (ppm) | σ | Pb (ppm) | σ | Sn (ppm) | σ | Tl (ppb) | σ | Bi (ppb) | σ | In (ppb) | σ |
|---|---|---|---|---|---|---|---|---|---|---|---|---|---|---|---|---|
| ALH 83100 | 13.9 | 0.32 | 0.10 | 3 | 127 | 14 | 0.92 | 0.01 | 0.45 | 0.01 | 50 | 7 | 28 | 5 | 29 | 3 |
| *ALH 84033* | 6.9 | 0.43 | 0.04 | 3 | 171 | 19 | 1.25 | 0.01 | 0.64 | 0.02 | 70 | 14 | 38 | 9 | 37 | 6 |
| LEW 85312 | 8.2 | 0.35 | 0.06 | 4 | 155 | 17 | 1.22 | 0.01 | 0.60 | 0.02 | 72 | 15 | 39 | 10 | 33 | 6 |
| LON 94101 | 10.6 | 0.32 | 0.01 | 3 | 144 | 17 | 1.31 | 0.01 | 0.69 | 0.03 | 78 | 21 | 42 | 14 | 36 | 9 |
| *MIL 05152* | 5.5[c] | 0.45 | 0.07 | 3 | 161 | 18 | 1.18 | 0.18 | 0.66 | 0.07 | 69 | 18 | 38 | 11 | 39 | 7 |
| MIL 07700 | 5.5 | 0.31 | 0.07 | 3 | 158 | 18 | 1.04 | 0.01 | 0.56 | 0.01 | 53 | 11 | 28 | 7 | 33 | 4 |
| *PCA 02010* | 1.1 | 5.42 | 0.44 | 2 | 1 | 0.3 | 0.06 | 0.01 | 0.00 | 0.01 | 0 | 5 | 1 | 5 | 0 | 6 |
| *PCA 02012* | 1.7 | 16.73 | 0.09 | 4 | 12 | 1 | 0.17 | 0.01 | 0.10 | 0.01 | 0 | 5 | 2 | 5 | 0 | 6 |
| EET 83355 | 3.6 | 0.36 | 0.06 | 3 | 85 | 9 | 0.63 | 0.01 | 0.29 | 0.02 | 43 | 15 | 14 | 10 | 15 | 6 |
| BUC 10933 | - - - | 0.26 | 0.06 | 3 | 80 | 9 | 0.64 | 0.01 | 0.30 | 0.02 | 34 | 14 | 17 | 9 | 12 | 6 |
| *GRA 06100* | 0.00 | -1.02 | 0.08 | 3 | 63 | 9 | 7.58 | 2.93 | - - - | - - - | - - - | - - - | - - - | - - - | - - - | - - - |
| *GRO 03116* | 1.2 | 4.77 | 0.03 | 3 | 43 | 6 | 0.17 | 0.03 | - - - | - - - | - - - | - - - | - - - | - - - | - - - | - - - |
| GRO 95577 | 12.3 | 0.33 | 0.02 | 3 | 203 | 23 | 1.68 | 0.01 | 0.69 | 0.02 | 74 | 16 | 64 | 11 | 48 | 6 |
| LAP 02342 | 3.1[c] | 0.09 | 0.01 | 3 | 39 | 4 | 0.27 | 0.00 | 0.12 | 0.01 | 18 | 11 | 6 | 7 | 5 | 4 |
| MET 00426 | 0.36 | 0.13 | 0.04 | 3 | 44 | 7 | 0.42 | 0.08 | - - - | - - - | - - - | - - - | - - - | - - - | - - - | - - - |
| *RBT 04133* | 1.5 | 0.11 | 0.04 | 3 | 78 | 9 | 0.58 | 0.01 | 0.27 | 0.01 | 32 | 11 | 20 | 8 | 14 | 4 |
| BHVO-2 | | 0.32[d] | 0.02 | - - - | 106 (101)[e] | 13 | 1.49 (1.51)[e] | 0.14 | 2.19 (1.7)[f] | 0.01 | - - - | - - - | - - - | - - - | - - - | - - - |
| CI[g] | ~ 20[h] | | | | 315 | | 2.50 | | 1.70 | | 142 | | 110 | | 80 | |
| CM[g] | ~ 9[h] | | | | 180 | | 1.6 | | 0.790 | | 92 | | 71 | | 50 | |
| CR[g] | < 9[h] | | | | 100 | | - - - | | 0.730 | | 60 | | 40 | | 30 | |
| T50 (K)[i] | | | | | 726 | | 727 | | 704 | | 532 | | 746 | | 536 | |

[a] Samples in *italics* are those identified as heated by: Alexander et al. (2012) for CMs, Abreu (2011) for GRA 06100 and GRO 03116, Davidson et al. (2014) for RBT 04133.
[b] Water loss as reported in Table 1 (Garenne et al., 2014).
[c] Calculated stoichiometrically from H contents (Alexander et al., 2012) and the 1/8 mass ratio of H to O in water.
[d] Mahan et al. (2017).
[e] Values in ( ) from Barrat et al. (2012).
[f] Value in ( ) from Gale et al. (2013).
[g] Average values reported in Fegley and Lodders (1998).
[h] Estimates reported in Garenne et al. (2014) and references therein.
[i] Reported values in Lodders (2003).

Table 3 – Trace element concentrations from literature used in this study

| Source | Sample[a] | Type[b] | Weathering[b] | Zn (ppm) | Tl (ppb) | Bi (ppb) | In (ppb) |
|---|---|---|---|---|---|---|---|
| *Paul and Lipschutz 1989* | Y 82042 | CM1/2 | - - - | 196 | 85 | 68 | 48 |
| *Xiao and Lipschutz, 1992* | ALH A81002 | CM2 | Ae | 183 | 75 | 62 | 47 |
| | ALH 84039 | CM2 | A/B | 193 | 85 | 83 | 52 |
| | ALH 83100 | CM1/2 | Be | 132 | 70 | 53 | 43 |
| | *Y 793321* | CM2 | - - - | 166 | 92 | 72 | 38 |
| | *MAC 88100* | CM2 | Be | 160 | 83 | 68 | 50 |
| *Zolensky et al., 1992* | Murchison | CM2 | - - - | 164 | 90 | 73 | 42 |
| *Zolensky et al., 1997* | Cold Bokkeveld | CM2 | - - - | 209 | 93 | 78 | 48 |
| | LEW 90500 | CM2 | B | 166 | 84 | 51 | 47 |
| *Wang and Lipschutz 1998* | *A 881655* | CM2 | - - - | 98 | 49 | 43 | 29 |
| | *Y 791198* | CM2 | - - - | 162 | 82 | 75 | 55 |
| | *Y 793321* | CM2 | - - - | 230 | 80 | 59 | 44 |
| | *Y 86789* | C2-ungrouped | - - - | 32 | 6 | 11 | 12 |

[a] Samples in *italics* are those identified as heated in their respective source studies (and references therein).
[b] Petrologic type and weather grade from Meteoritical Bulletin (where available).

Table 4 – Volatility sequences for CM (and CR) carbonaceous chondrites

| Source | Type | Increasing volatility → |
|---|---|---|
| *This study* | CM | Pb, Sn, Bi, In, Zn, Tl |
|  | CR | Tl, Zn, Sn, Pb, Bi, In |
| *Matza & Lipschutz (1977)* | CM | Zn, In, Bi, Tl |
| *Lodders (2003)* | CM | Bi, Pb, Zn, Sn, In, Tl |

Figure 1

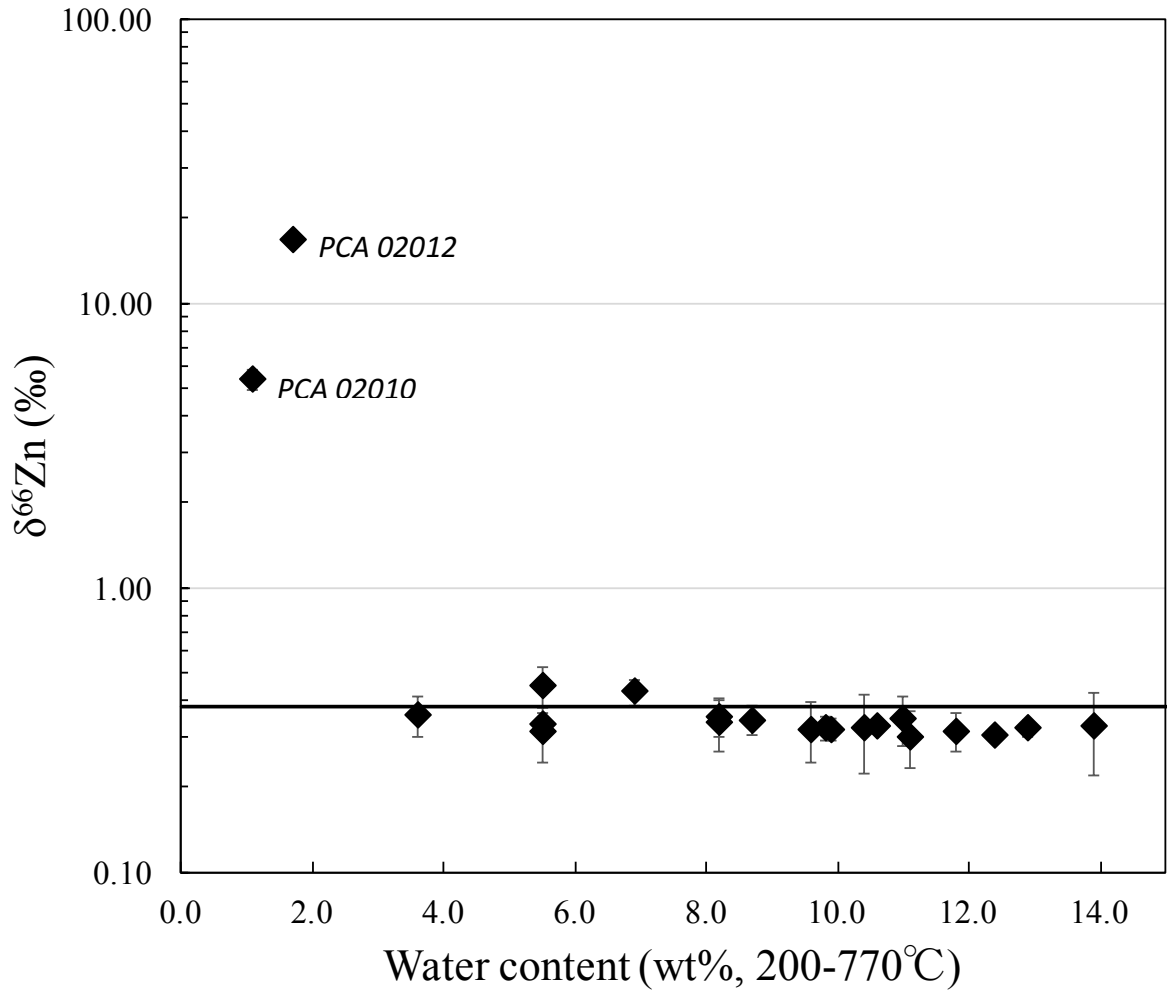

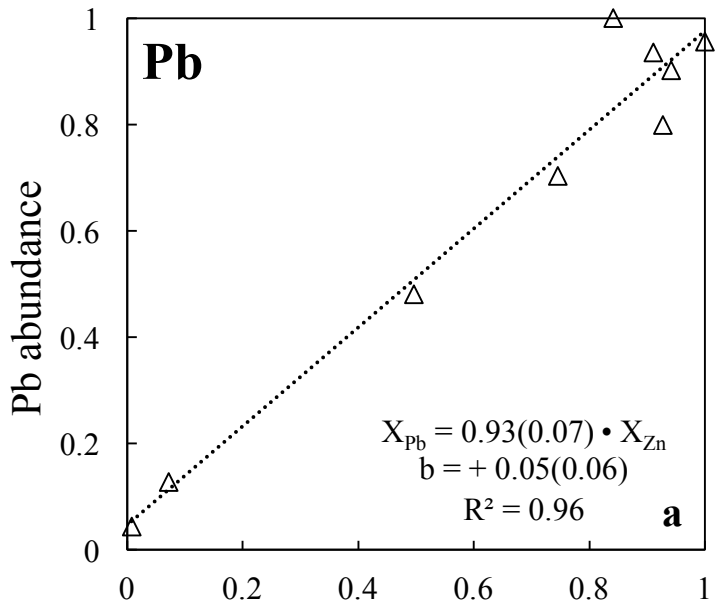
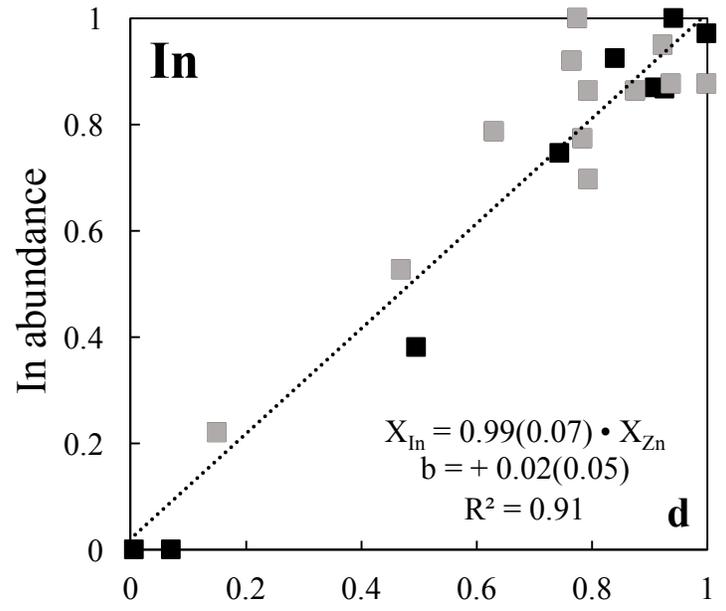
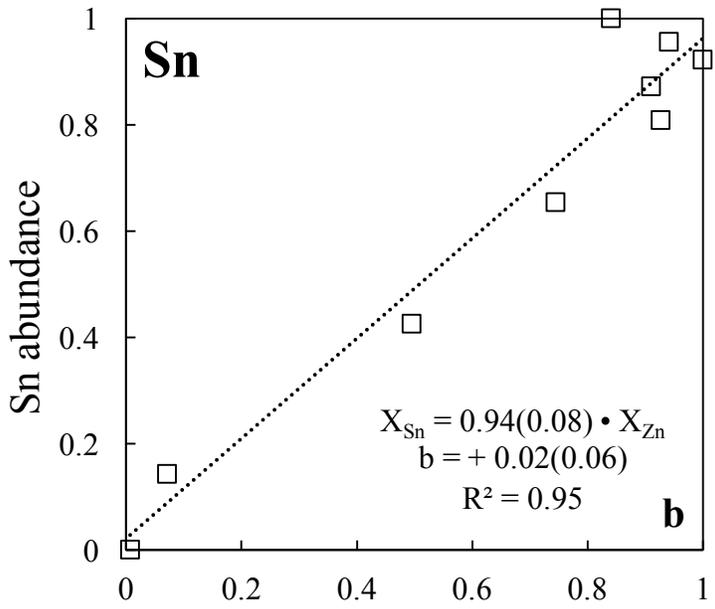
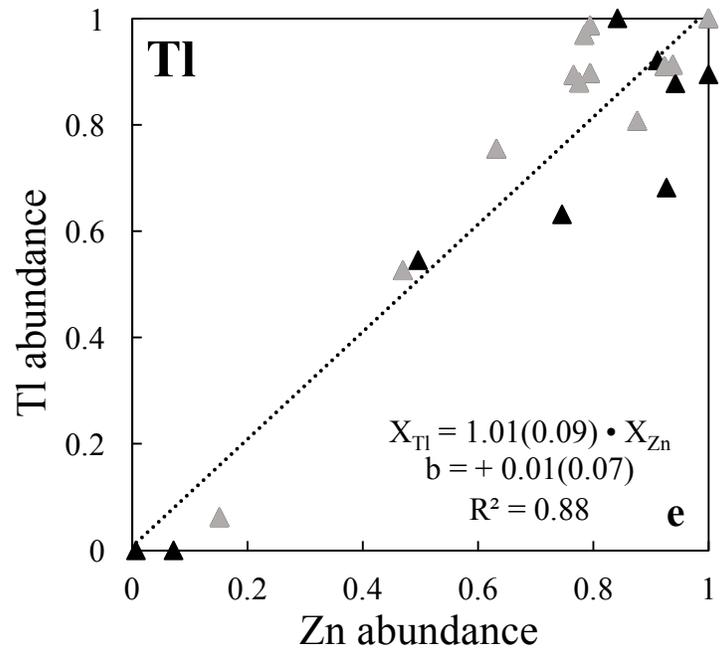
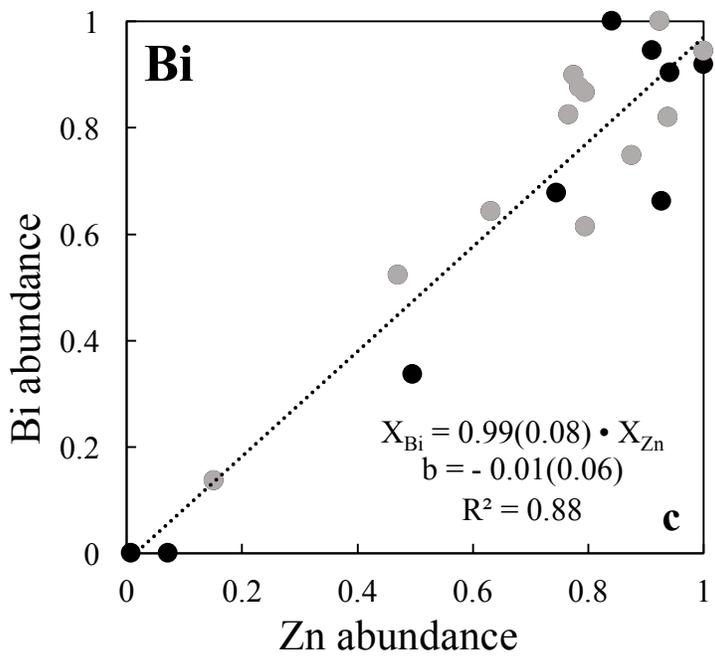
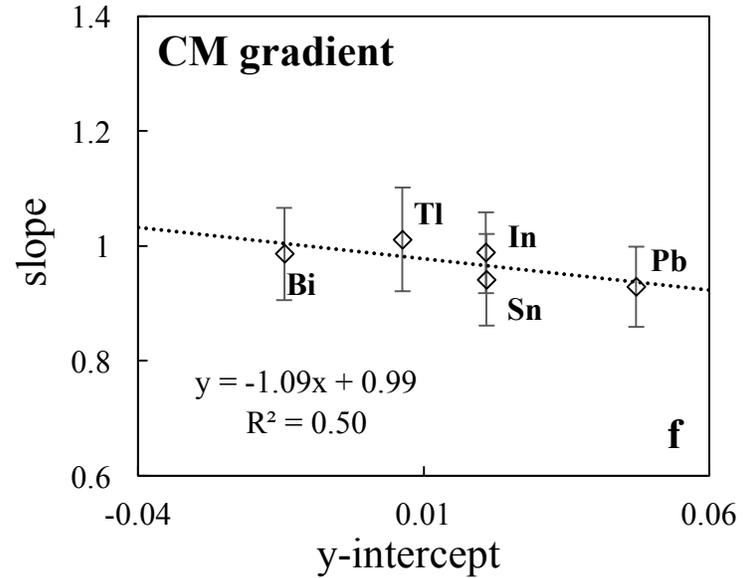

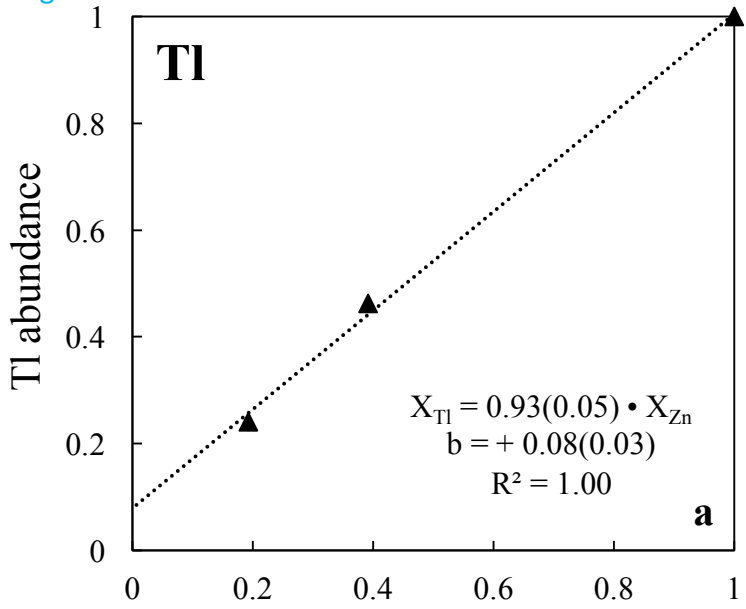
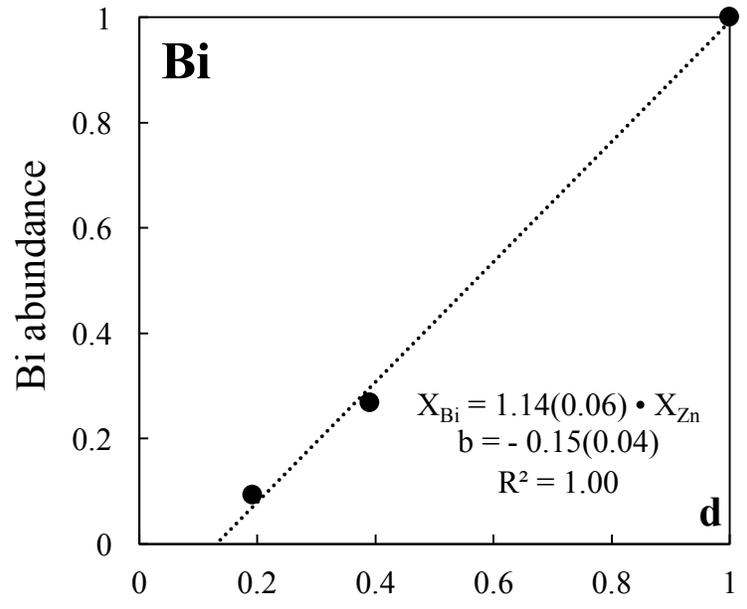
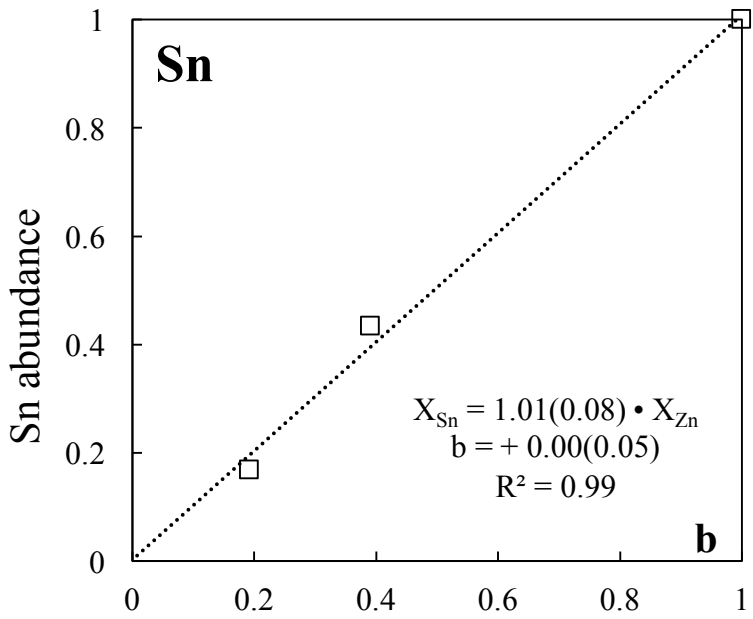
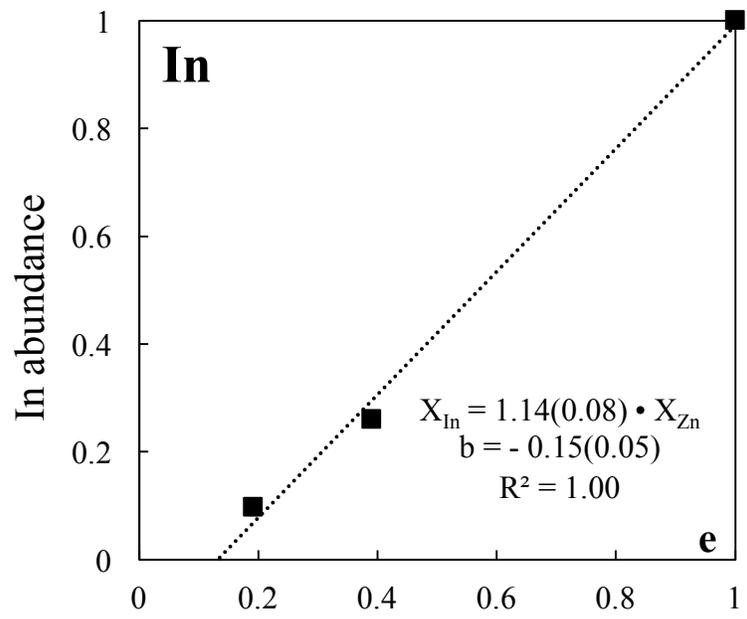
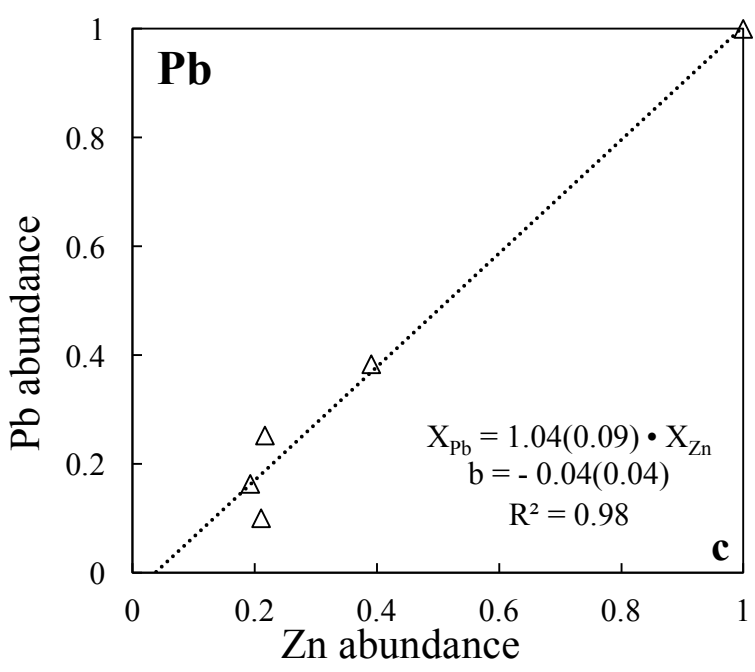
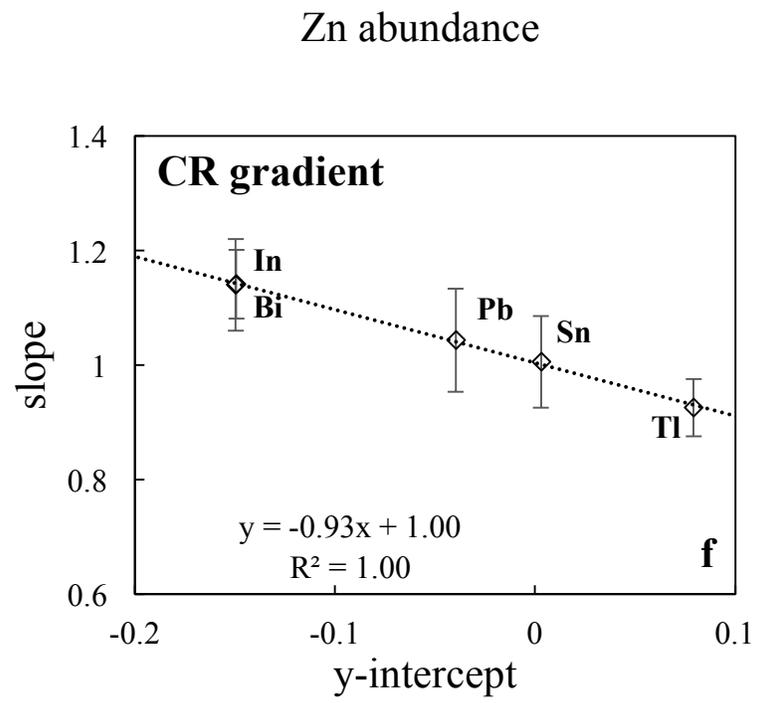

Figure 3

Figure 4

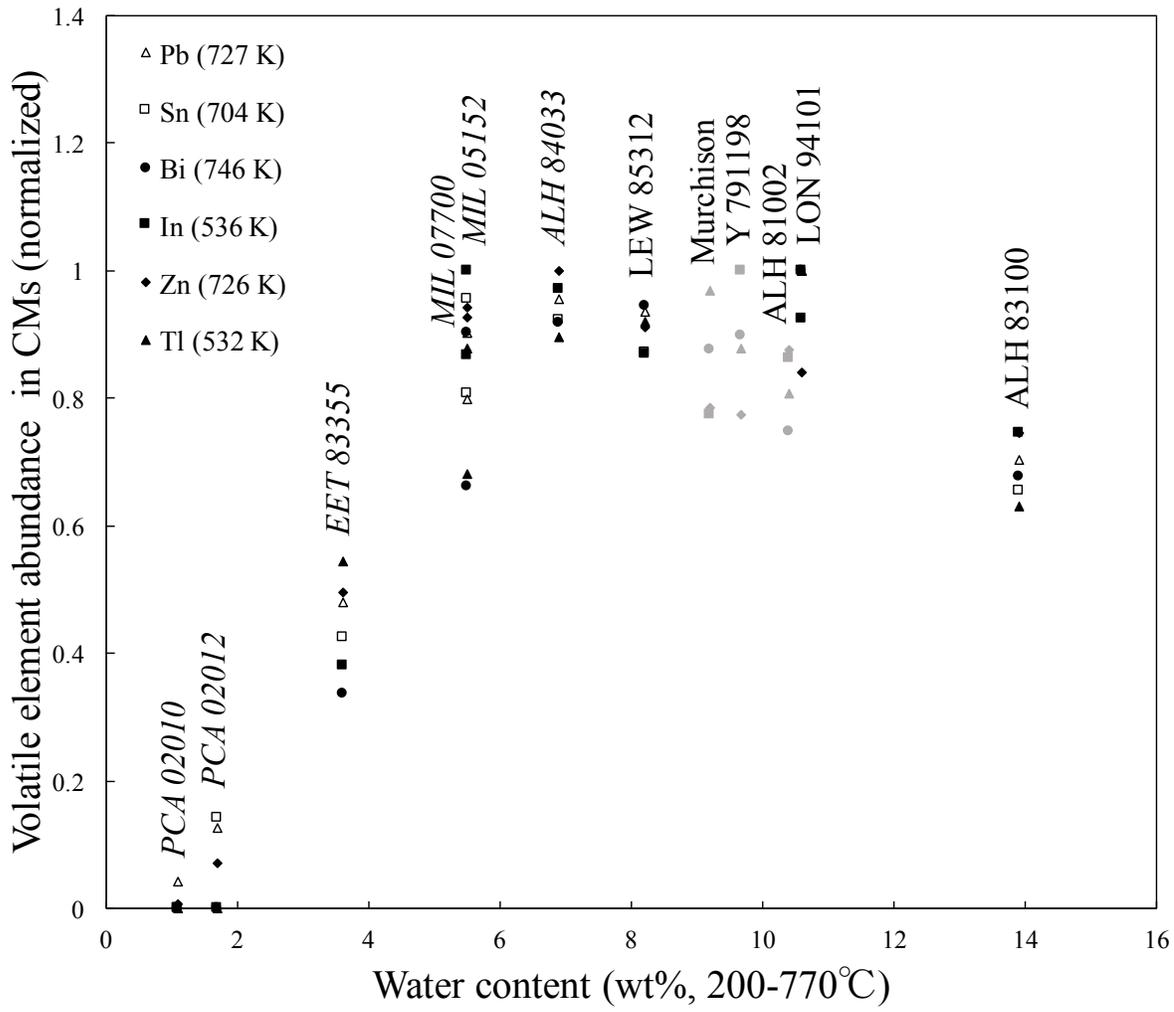

Figure 5

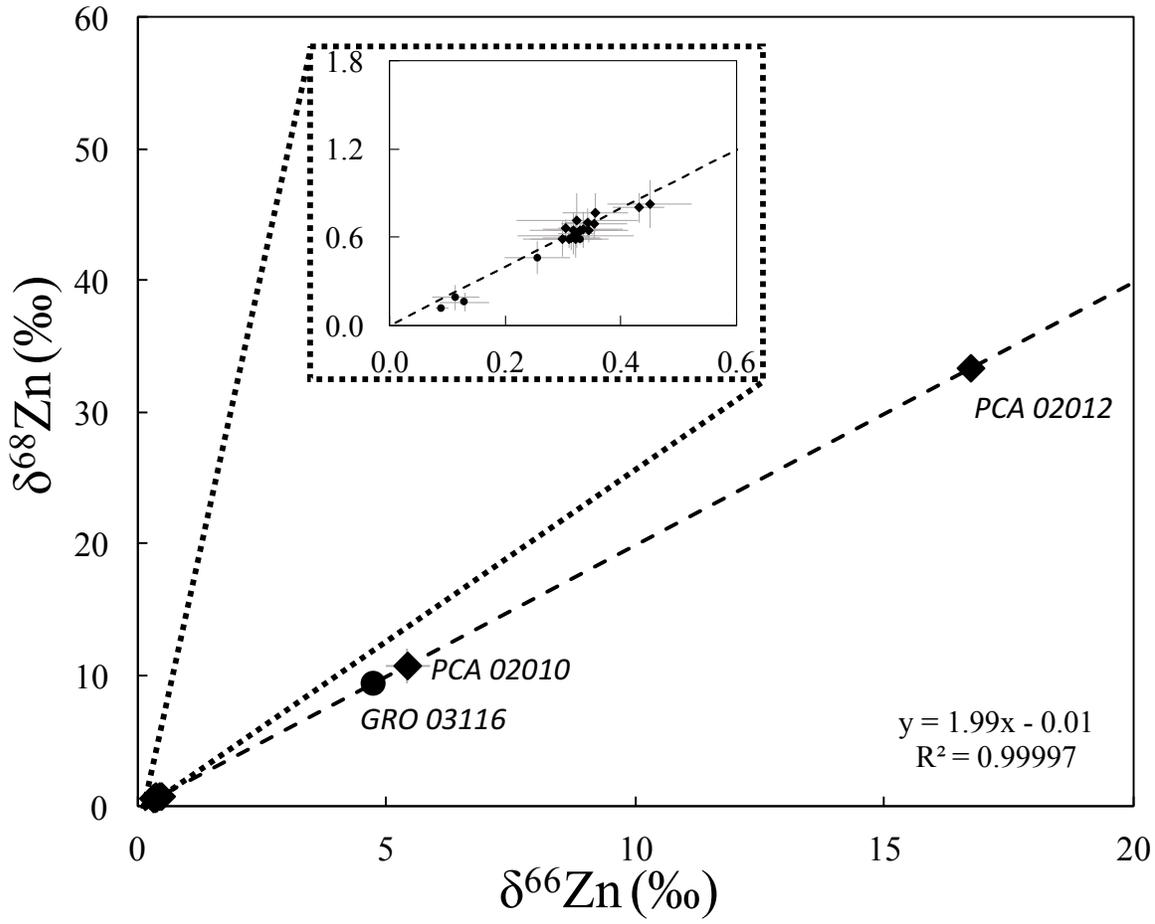